\documentclass[11pt,a4paper]{article}
\usepackage{jheppub,amsmath, amsthm, amssymb,slashed,url}
\usepackage{graphicx}
\usepackage{epstopdf}
\def\t{{ \sf t}}

\def\be{\begin{equation}}
\def\ee{\end{equation}}

\def\cs{{\mathrm{cs}}}
\def\hat{\widehat}
\def\tilde{\widetilde}

\def\D{{\mathcal D}}
\def\S{{\mathcal S}}
\def\V{{\mathcal V}}

\def\SIgma{\Sigma}

\def\red{{\mathrm{red}}}

\def\d{{\mathrm d}}

\def\R{{\mathbb R}}
\def\C{{\mathbb C}}

\def\D{{\mathcal D}}

\def\G{{\mathcal G}}
\def\[{\bigl [}
\def\]{\bigr ]}
\def\CP{{\mathbb{CP}}}

\def\Z{{\mathbb Z}}

\def\LL{{ \mathfrak L}}
\def\t{\tilde }
\def\h{\widehat}

\def\V{{\mathcal V}}

\def\NS5{\mathrm{NS5}}
\def\uD5{\mathrm{D5}}

\def\l{\langle}
\def\r{\rangle}

\def\epsilon{\varepsilon}

\def\e{{\mathbf e}}
\def\i{{\mathbf i}}
\def\spin{{\mathrm{spin}}}

\def\tilde{\widetilde}
\def\bar{\overline}
\def\neg{\negthickspace}
\def\Ber{{\mathrm {Ber}}}
\font\teneurm=eurm10 \font\seveneurm=eurm7 \font\fiveeurm=eurm5
\newfam\eurmfam
\textfont\eurmfam=\teneurm \scriptfont\eurmfam=\seveneurm
\scriptscriptfont\eurmfam=\fiveeurm

\font\teneusm=eusm10 \font\seveneusm=eusm7 \font\fiveeusm=eusm5
\newfam\eusmfam
\textfont\eusmfam=\teneusm \scriptfont\eusmfam=\seveneusm
\scriptscriptfont\eusmfam=\fiveeusm

\font\tencmmib=cmmib10 \skewchar\tencmmib='177
\font\sevencmmib=cmmib7 \skewchar\sevencmmib='177
\font\fivecmmib=cmmib5 \skewchar\fivecmmib='177
\newfam\cmmibfam
\textfont\cmmibfam=\tencmmib \scriptfont\cmmibfam=\sevencmmib
\scriptscriptfont\cmmibfam=\fivecmmib

\def\M{{\mathcal M}}
\def\MM{{\mathfrak M}}
\def\Pi{\varPi}

\title{Notes On Supermanifolds And Integration}

 \author{Edward Witten}
\affiliation{School of Natural Sciences, Institute for Advanced Study,\\ 1 Einstein Drive, Princeton, NJ 08540 USA}
\abstract{These are notes on the theory of supermanifolds and integration on them, aiming
to collect results that are useful for a better understanding of superstring perturbation theory in the RNS
formalism.  
}
\begin{document} \maketitle

\section{Introduction}\label{intro}

Supersymmetric field theories have been studied from many points of view since their discovery roughly forty years ago.
Formulating a supersymmetric field theory in superspace -- that is on a supermanifold -- is, when possible,  often very helpful.
In practice, however, natural physics questions often require only the most basic facts about supermanifolds.

One topic  stands out as a conspicuous exception.  This is superstring perturbation theory in the RNS formalism.  This perturbation
theory is formulated in terms of integration on the moduli space of super Riemann surfaces.  That moduli space is a rather subtle supermanifold
and simple questions about superstring perturbation theory quickly lead to relatively subtle issues of supergeometry.  Superstring perturbation theory
really does seem like one topic that can be better understood with more input from supergeometry.

The present notes aim to present background material on supermanifolds and integration.
The material is not novel, except possibly for a few details, and the presentation does not aim for either completeness or full rigor.  Rather, the goal has been to collect in 
a relatively simple way some background material for a reconsideration of superstring perturbation theory, which  will appear 
elsewhere \cite{reconsidered}.    A companion article will contain background
material on super Riemann surfaces \cite{supersurf}.  

Of course, there is an extensive literature on this topic and it is impossible to give complete references.  
Much of the material outlined here can be found in books such as \cite{Berezin,Manin,DeWitt,Rogers}
and review articles such as \cite{Kac,Leites,Voronov,Nelson}.  A useful and extremely concise introduction is \cite{DEF}.
  The fundamental 
structure theorem for smooth supermanifolds was proved in \cite{Ber,Bach,Gaw} and
the theory of integral forms was initiated in \cite{BL}.  Other useful references include \cite{BaSch,VZ,VZtwo,GMM}.
The superstring literature is likewise too vast to be cited in full.   The classic work \cite{FMS} introduced some key concepts such as the role of different
representations of the Weyl algebra, the papers \cite{MNP,DPh,superoperator,EHVerl} construct measures on supermoduli space via superconformal field theory, and
the paper \cite{RSV}, which is unfortunately little-known, does this via algebraic geometry.  The papers \cite{Bel,Beltwo,Belthree}, which again are unfortunately little-known, 
are valuable both as an exposition of aspects of supergeometry
and for insight about its role in superstring perturbation theory.

In section \ref{supermanifolds}, we describe the basic idea of a supermanifold.  In section \ref{superintegration}, we sketch the theory of integration on supermanifolds, and 
in section \ref{operations}, we describe some additional useful facts and constructions.  Section \ref{bicomplex} is devoted
to a close look at some basic ideas needed in string perturbation theory.

\section{Supermanifolds}\label{supermanifolds}

\subsection{Smooth Supermanifolds}\label{smooth}

Roughly speaking, a supermanifold $M$ of dimension $p|q$ (that is, bosonic dimension $p$ and fermionic dimension $q$) can be described
locally by $p$ bosonic coordinates $t^1\dots t^p$ and $q$ fermionic coordinates $\theta^1\dots\theta^q$.  Sometimes we abbreviate the whole
collection of coordinates as $t^1\dots|\dots \theta^q$ or simply as $x$.  

We cover $M$ by open sets $U_\alpha$ each of which  can be described by coordinates $t^1_\alpha\dots |\dots \theta^q_\alpha$.
On the intersection $U_\alpha\cap U_\beta$, the $t^i_\alpha$ are even functions of $t^1_\beta\dots|\dots\theta^q_\beta$ and the $\theta^s_\alpha$
are odd functions of the same variables.  We call these functions gluing functions and denote them as $f_{\alpha\beta}$ and $\psi_{\alpha\beta}$:
\begin{align}\label{orro} t^i_\alpha & = f_{\alpha\beta}^i(t^1_\beta\dots|\dots \theta^q_\beta) \cr
                                                      \theta^s_\alpha& = \psi_{\alpha\beta}^s(t^1_\beta\dots|\dots\theta^q_\beta).\end{align}
On  the intersection $U_\alpha\cap U_\beta$, we require that the gluing map defined by $f_{\alpha\beta}^1\dots|\dots\psi^q_{\alpha\beta}$ is inverse to the 
one defined by $f_{\beta\alpha}^1\dots|\dots\psi_{\beta\alpha}^q$,
and we require an obvious compatibility of the gluing maps on triple intersections $U_\alpha\cap U_\beta\cap U_\gamma$.

Now we have to be more precise about what sort of supermanifold we want.  The most obvious notion is a real supermanifold of dimension $p|q$.
This would mean that the $t^i_\alpha$ and $\theta^s_\alpha$ are all real variables, and the gluing functions $f^i_{\alpha\beta}$ and $\psi^s_{\alpha\beta}$ are
all real.  To be more pedantic, reality of the gluing functions means that if we expand these functions in powers of the $\theta$'s, for example
\begin{equation}\label{expo}f^i_{\alpha\beta}(t\dots|\dots\theta)=g^i_{\alpha\beta}(t^1_\beta\dots t^p_\beta)+\sum_s\theta^s_\beta g^i_{\alpha\beta s}(t^1_\beta\dots t^p_\beta)+\dots ,\end{equation}
then, for real $t^i_\beta$, the functions arising in this expansion are all real.\footnote{\label{dolf}For the moment we consider a single supermanifold $M$ rather than a family of supermanifolds
parametrized by some other space, so we assume that the gluing functions depend only on $t^1\dots|\dots\theta^q$.  See section
\ref{families}.}  
If this condition is obeyed, we say that $M$ is a real supermanifold and that for each $\alpha$ the
local coordinate system $t^1_\alpha\dots|\dots\theta^q_\alpha$ gives an isomorphism of  $U_\alpha$ with an open set in $\R^{p|q}$.

Real supermanifolds are  the right framework for superspace descriptions of supersymmetric field theories in Lorentz signature and for many
other applications in Lorentz signature. But
they are often not convenient for Euclidean signature quantum field theory, largely because spinors in Euclidean signature often do not admit a real structure. A  related fact is that they are not 
 convenient for superstring perturbation theory.  
 The most important supermanifolds for superstring perturbation theory are super Riemann surfaces and the moduli spaces thereof;  in each case, the fermionic variables have
no real structure, so these are not real supermanifolds.

 For superstring perturbation theory and for many other Euclidean signature applications, one wants a more general notion that is called a cs manifold  in \cite{DEF},
p. 94 (where it is stated that cs stands for complex supersymmetric).  
Informally, in  a cs manifold, although the $t$'s are real at $\theta=0$, there is no reality condition (for either $t$'s or $\theta$'s) for $\theta\not=0$.  
To be more precise, in terms of the gluing functions, we require that the bosonic gluing
functions $f^i_{\alpha\beta}$ are real at $\theta^1=\dots=\theta^q=0$, but we impose no reality condition on the $\theta$-dependent terms in $f^i_{\alpha\beta}$, and no
reality condition at all on $\psi^i_{\alpha\beta}$.
   In the case of a cs manifold, we say that the coordinate functions $t^1_\alpha\dots|\dots\theta^q_\alpha$
give an isomorphism of the set $U_\alpha$ with an open set in $\R^{p|*q}$, where the asterisk is meant to remind us that  there is only
a real structure when the odd variables vanish.    In this
paper, when not stated otherwise, our ``supermanifolds'' are  cs manifolds.  It is usually clear that the statements can be naturally specialized to real supermanifolds.
On rare occasions, we note differences between the two cases. 

On a cs supermanifold, there is no notion of taking the complex conjugate of a function.
This only makes sense once the odd variables are set to zero.   In particular, we are never allowed to talk about $\bar \theta$, a hypothetical
complex conjugate of an odd variable $\theta$.

An important point  is that to a supermanifold $M$, one can in a natural way associate a reduced space $M_\red$ that is an ordinary real manifold, naturally embedded in $M$, and of the
same bosonic dimension.
One simply sets the odd variables $\theta^1_\alpha\dots\theta^q_\alpha$ to zero in the gluing law.  This is consistent because the odd gluing functions $\psi^s_{\alpha\beta}$
are of odd order\footnote{We still make the assumption of footnote \ref{dolf}, so the only odd variables that can appear in the gluing functions
are $\theta^1\dots\theta^q$.} in $\theta^1_\beta\dots\theta^s_\beta$ and hence vanish
when the $\theta$'s do, so the gluing law implies that all 
$\theta^i_\alpha$ vanish if and only if all $\theta^i_\beta$ do.  Moreover, once we set the $\theta$'s to zero, the $f^i_{\alpha\beta}$ become real, by the definition of a cs manifold.  
The functions $f^i_{\alpha\beta}(t^1_\beta\dots t^p_\beta|0\dots0)$ are then the gluing functions of an ordinary $p$-dimensional manifold that we call $M_\red$.  Moreover, there is
a natural embedding
\begin{equation}\label{natemb}i:M_\red\to M\end{equation}
that takes the point in $M_\red$ with coordinates $t^1_\alpha\dots t^p_\alpha$ to the point in $M$ with coordinates $t^1_\alpha\dots|0\dots 0$.

Though we have defined supermanifolds by means of gluing, they can also be defined by any familiar method for defining ordinary manifolds.  
For instance, a real supermanifold $M$ of dimension $2|2$ can be defined by a real equation such as
\begin{equation}\label{zobo} x^4+y^4+z^4+\theta_1\theta_2 =1,\end{equation}
with real variables $x,y,z,\theta_1,\theta_2$.   To present $M$ in the gluing language, one would for example cover it by open sets $U_\alpha$ in each of which one can solve
for one of the bosonic coordinates $x,y, $ or $z$ in terms of the other variables (for example, in one open set, one might solve for $z$ by  $z=(1-x^4-y^4-\theta_1\theta_2)^{1/4}$).
The equation (\ref{zobo}) defines a real supermanifold because the parameters in the equation are all real.  To get a cs manifold that is not real, one could
add to the equation an additional term that is not real but that vanishes at $\theta_1=\theta_2=0$.  For example, if $\lambda$ is a complex number that is not real, then a suitable equation is
\begin{equation}\label{obo}x^4+y^4+z^4+\theta_1\theta_2(1+\lambda x^2)=1.\end{equation}

\subsubsection{Families of Supermanifolds}\label{families}

Often one wishes to consider not a single supermanifold $M$ but a family of supermanifolds parametrized by some other supermanifold $N$.  For example,
$M$ might be a super Riemann surface, which depends on bosonic and fermionic moduli that parametrize $N$; in this example, $N$  could be the moduli
space of super Riemann surfaces.  The best way to think about this situation is to consider a supermanifold $X$ that is fibered over $N$ with the fibers being copies
of $M$.

In this situation, $X$ is a supermanifold in the sense that we have already described and therefore it has a reduced space $X_\red$.   In defining $X_\red$,
all odd variables are set to zero, both the odd parameters in $N$, which we will call $\eta^1\dots\eta^s$, and the odd parameters $\theta^1\dots\theta^q$
in $M$.

Though $M$ depends on $\eta^1\dots\eta^s$, it does not have a reduced space that depends on those parameters.  The reason is
that since the gluing functions $\psi^i_{\alpha\beta}$ can depend on the $\eta$'s, we will in general get gluing laws such as $\theta_\alpha=\theta_\beta+\eta$ and
we cannot consistently set the $\theta$'s to zero unless we also set the $\eta$'s to zero.

So for example if $M$ is a single super Riemann surface, it has a reduced space $M_\red$ that is an ordinary Riemann surface.  But if $M$ depends on some odd parameters
$\eta_1,\dots,\eta_s$, then we cannot define a reduced space without setting those parameters to zero.  That is why there is no elementary map from the moduli space
of super Riemann surfaces to the moduli space of ordinary Riemann surfaces.  This fact led to complications in the superstring literature of the 1980's.

\subsubsection{Open Sets And Other Topological Notions}\label{opensets}

The intuition concerning the concept of an ``open set'' is that if $U$ is an 
open set in a topological space $Y$, and $p$ is a point in $U$, then any point in $Y$
 sufficiently close to $p$ is also contained
in $U$.  

Now suppose that $y$ is an even coordinate on a supermanifold $M$ and 
that, for example, $\zeta_1$ and $\zeta_2$ are odd quantities (either odd 
functions on $M$ or odd moduli).
Then since $\zeta_1\zeta_2$ is nilpotent, we should think of 
it as smaller in value than any complex number.  So we should consider, 
for instance, $y+\zeta_1\zeta_2$ to
be ``sufficiently close'' to $y$ in any reasonable sense.

The upshot of this is that it is not helpful to introduce a concept of 
``open set in $M$'' that is different from the concept ``open set in $M_\red$.''    
We just consider the two
concepts to be synonyms. So similarly the statement ``the $U_\alpha$ form an 
open cover of $M$'' means the same as ``the $U_\alpha$ form an open cover 
of $M_\red$.''  For another example, a ``neighborhood'' in $M$ of a subset 
$Y_\red\subset M_\red$ is an open set in $M$ whose reduced 
space is a neighborhood of $Y_\red$ in $M_\red$.

A common approach in rigorous mathematical treatments is to define the 
$U_\alpha$ as open sets in $M_\red$, but to endow each $U_\alpha$
with a larger ring of functions than the obvious ring of smooth 
functions on $U_\alpha$ -- namely the functions of the whole set of 
even and odd coordinates $t^1\dots|\dots\theta^q$,
with the $t$'s restricted to $U_\alpha$.  (In a more fancy language, $M_\red$ is endowed
with ``a sheaf of Grassmann algebras.'') 
We will not need this language in these notes.  
However, the reader might find it helpful to develop the intuition that because fermions
are infinitesimal, covering $M$ by open sets is equivalent to covering $M_\red$ by open sets.

More generally, for similar reasons, one identifies various topological 
notions on $M$ with the same notions for $M_\red$.
For example, an orientation or spin structure on $M$ is by definition an 
orientation or spin structure on $M_\red$.  One says that $M$ is compact
if and only if $M_\red$ is compact.  
The Euler characteristic of $M$ is defined to be that of $M_\red$, and 
if $M_\red$ is a Riemann surface of genus $g$,
then we also refer to $g$ as the genus of $M$.

\subsection{Submanifolds Of A Smooth Supermanifold}\label{submanifolds}

We will now describe another general fact about a smooth supermanifold $M$.  If $N_\red$ is any submanifold of $M_\red$ of codimension $r$, then one can ``thicken'' it slightly in the fermionic
dimensions to make a submanifold\footnote{We usually refer to sub-supermanifolds simply as submanifolds, as the term sub-supermanifold is clumsy.}
$N\subset M$  of codimension $r|0$.  $N$ is not unique, but it is unique up to ``homology,'' in fact 
up to an 
infinitesimal wiggling in the odd directions.  The importance of this is that once we know the 
appropriate analog of a differential form on a supermanifold, a closed form on $M$ of the appropriate degree
(for example, an integral form of codimension $r$ in the language of section \ref{informs}) can be integrated over $N$ with a result that does not depend on the choice of thickening.  In addition to being uniquely determined up to homology, 
$N$ is also uniquely determined as a supermanifold, up to a (non-unique) diffeomorphism that acts trivially on $N_\red$.

One should think of existence of $N$ 
as an intuitively obvious reflection of the fact that the fermionic directions in a supermanifold
are infinitesimal.  Roughly, if $N_\red\subset M_\red$ is defined
locally by the vanishing of $r$ real-valued functions $h_1,\dots,h_r$, then upon taking the $h$'s to depend upon the odd coordinates $\theta^s$  in an arbitrary fashion, one
gets functions on $M$ whose vanishing defines the desired submanifold $N$.  A precise proof follows from what one may call the fundamental structure theorem of smooth supermanifolds.   As described 
below, this structure theorem lets one construct a (non-unique) projection $\pi:M\to M_\red$, and one can define  $N$ by $N=\pi^{-1}(N_\red)$. 
For  the original proofs of the structure theorem, see \cite{Ber, Bach, Gaw}; see for example section 4.2 of  \cite{Manin},  section 3 of   \cite{Voronov}, or Theorem 8.2.1 of \cite{Rogers} for expositions.

To explain the structure theorem, observe first that instead of merely setting the $\theta$'s to zero in the gluing relations, we could consistently drop
all terms of order $\theta^2$ and higher.  This puts the gluing relations in the following form:
\begin{align}\label{zoro}    t^i_\alpha & = f_{\alpha\beta}^i(t^1_\beta\dots t^p_\beta) \cr
                                                      \theta^s_\alpha& = \sum_u\psi_{\alpha\beta\,u}^s(t^1_\beta\dots t^p_\beta)\theta^u_\beta.\end{align}     
Here we should think of the objects $\psi_{\alpha\beta\,u}^s$ as matrix elements of a linear transformation $\psi_{\alpha\beta}$ acting on 
the odd variables.  The consistency relations on the gluing data say that these linear transformations are the transition matrices of
a vector bundle $V\to M_\red$.  This is a bundle with purely odd fibers, of dimension $0|q$.

Thus, from every supermanifold $M$, one can extract an ordinary manifold $M_\red$ and a purely fermionic vector bundle\footnote{\label{humbug} The structure group
of this bundle is $GL(q,\C)$ in general.  If it cannot be reduced to $GL(q,\R)$ -- for example, the Chern classes of $V$ may present an obstruction --
then topologically the $\theta$'s cannot be given a real structure globally. In this case, $M$ must be viewed as a cs manifold rather than a real
supermanifold.  There is  never any
problem locally in giving the $\theta$'s a real structure, by picking a local basis
of $\theta$'s and declaring them to be real. In applications to string theory, there is typically a global obstruction to giving the $\theta$'s a real structure and no natural
way to do so locally.}  $V\to M_\red$.
The total space of this bundle is a supermanifold $M'$.  

The fundamental structure theorem says that as a smooth supermanifold, $M$ is always isomorphic to $M'$.  The proof is made by expanding
the gluing functions in a power series in the $\theta$'s and showing that, order by order, each term beyond those that we have kept
in (\ref{zoro}) can be eliminated by a suitable redefinition of the coordinates. (Moreover, the coordinate change in question shifts the $t$'s
only by terms of order $\theta^2$ or higher, and the $\theta$'s only by terms of order $\theta^3$ or higher.) 
Since there are only finitely many $\theta$'s, this process terminates
after finitely many steps.\footnote{Concretely, to eliminate the unwanted terms from the gluing functions, one needs
to know vanishing of certain sheaf cohomology classes on $M_\red$ that can be extracted from the expansion of the gluing functions in powers of
$\theta$.   These classes all vanish because in general  sheaf cohomology of any smooth
manifold, such as $M_\red$, with values in the sheaf of sections of any vector bundle always vanishes except in degree zero.  
The analogous sheaf cohomology for complex manifolds is in general
nonzero, which is why a complex supermanifold need not be holomorphically split.  See section \ref{omgo}.}

A supermanifold that is presented in the form (\ref{zoro})  is said to be split.   The structure theorem says that every smooth supermanifold
can be split, but not in a unique fashion. 
Once a splitting is picked, there is a natural  projection map $\pi:M\to M_\red$  that simply forgets
the $\theta$'s (and thus maps  the point in $M$ labeled by $t^1_\alpha\dots|\dots \theta^q_\alpha$ to the point in 
$M_\red$ labeled by $t^1_\alpha\dots t^p_\alpha$).  
The projection $\pi$ is related to the inclusion $i:M_\red\to M$ that we defined earlier by $\pi\circ i = 1$.

The drawback of the structure theorem is that the projection $\pi:M\to M_\red$ whose existence is guaranteed by the theorem is far from unique,
and the theorem comes with no advice about finding a useful or natural choice.
 In the context of superstring perturbation theory, for example, the structure theorem says (modulo some issues
 discussed in section \ref{bicomplex}) that if we wish we can pick a projection
$\pi$ from the moduli space $\MM$ of super Riemann surfaces to the moduli space $\M$ of ordinary Riemann surfaces, and reduce a measure on $\MM$ to a measure
on $\M$ by integrating first over the fibers of this projection.  However, in the absence of a natural projection, this procedure may not be illuminating.

\subsection{Complex Supermanifolds}\label{complexsupermanifolds}

A complex supermanifold is defined similarly, except that the gluing functions are  holomorphic functions.  
$\C^{a|b}$ is a supermanifold parametrized by complex bosonic coordinates $z^1,\dots,z^a$ and fermionic coordinates $\theta^1\dots\theta^b$.
(Recall that in this paper there is never a reality condition on fermionic variables.) 
A complex supermanifold $ M$
of complex dimension $a|b$ can be covered by open sets $U_\alpha$ each of which is parametrized by even and odd complex coordinates $z^1_\alpha\dots z^a_\alpha$ and $\theta^1_\alpha\dots
\theta^b_\alpha$.  These coordinates give a holomorphic identification of $U_\alpha$ with an open set in $\C^{a|b}$.

On intersections $U_\alpha\cap U_\beta$ there are gluing relations analogous to  (\ref{orro}), with the difference
that the gluing functions are now required to be holomorphic in $z^1_\beta\dots|\dots\theta^b_\beta$:
\begin{align}\label{borro} z^i_\alpha & = f_{\alpha\beta}^i(z^1_\beta\dots|\dots \theta^b_\beta) \cr
                                                      \theta^s_\alpha& = \psi_{\alpha\beta}^s(z^1_\beta\dots|\dots\theta^b_\beta).\end{align}
To be precise, this holomorphy means that if the functions $f^i_{\alpha\beta}$ and $\psi^i_{\alpha\beta}$ are expanded as a polynomial
in the $\theta$'s, then the coefficient of each term is an ordinary holomorphic function of $z^1\dots z^a$.

A complex supermanifold $  M$ of dimension $a|b$ has a reduced space $M_\red$ obtained by setting the $\theta$'s to zero in the gluing relation.
The gluing relations then reduce to 
\begin{equation}\label{tono}z^i_\alpha=f^i_{\alpha\beta}(z^1_\beta\dots z^a_\beta|0\dots 0).\end{equation}
  These are gluing functions for an ordinary complex manifold
$M_\red$ of complex dimension $a$.  There is an evident holomorphic embedding $i:M_\red\to M$, mapping
$z^1_\alpha\dots z^a_\alpha$ to $z^1_\alpha\dots z^a_\alpha|0\dots 0$.
As in our discussion of (\ref{zoro}), by keeping in the gluing relations the terms that are linear in the $\theta$'s, we can define a holomorphic 
vector bundle $V\to M_\red$,
with fibers of dimension $0|q$.   The total space $M'$ of this bundle is a complex supermanifold that is an approximation to $M$ (but in contrast to the smooth case, $M'$ and $M$ are not necessarily isomorphic as complex supermanifolds; see section \ref{omgo}).

Examples of complex supermanifolds are easily given.  For example, let
us define complex projective superspace $\CP^{a|b}$ of dimension $a|b$.  It has homogeneous coordinates $z^1\dots z^{a+1}|\theta^1\dots\theta^b$,
subject to an overall scaling of all $z$'s and $\theta$'s by a nonzero complex parameter $\lambda$, and with a requirement that the bosonic coordinates
$z^a  $ are not allowed to all simultaneously vanish.  (In supermanifold theory, to say that a bosonic variable is ``non-zero'' means that it is invertible or in other
words remains nonzero after setting all odd variables to zero.)  
 To express $\CP^{a|b}$ in the above language, for $\alpha=1,\dots,a+1$, let $U_\alpha$ be defined by the condition
 $z^\alpha\not=0$. The $U_\alpha$ give an open cover of $\CP^{a|b}$.   Each $U_\alpha$ can be parametrized by the ratios $z^\beta/z^\alpha$, $\beta\not=\alpha$, as well as $\theta^j/z^\alpha$, with obvious holomorphic gluing relations.  

We can construct many additional examples  by imposing
an equation $F(z^1\dots |\dots\theta^b)=0$, where $F$ is a homogeneous polynomial in the homogeneous coordinates of $\CP^{a|b}$ that is either even or odd.  
If $F$ is sufficiently generic, this will give a complex supermanifold of dimension $a-1|b$ if $F$ is even, or of dimension $a|b-1$ if $F$ is odd.

\subsubsection{Holomorphic Splittings}\label{omgo}

We now consider for a complex supermanifold the questions analogous to those that for a smooth supermanifold
are addressed by the fundamental structure theorem described in section \ref{smooth}.

If  $M$ is a complex supermanifold, then the reduced space 
$M_\red$ is an ordinary complex manifold.  
We recall that there is always a natural holomorphic embedding $i:M_\red\to M$.
$M$ is said to be holomorphically projected if there is a holomorphic map $\pi:M\to M_\red$
with $\pi \circ i = 1$.  $M$ is said to be holomorphically split if the structure group of the fibration
$M\to M_\red$  given by $\pi$ reduces to the group $GL(q)$ of linear transformations of the fibers of $\pi$.
In other words, $M$ is holomorphically split if it is holomorphically isomorphic to $M'$, the total space of
a purely fermionic vector bundle $V\to M_\red$.

In terms of the gluing data (\ref{borro}), $M$ is holomorphically projected if the local coordinates can
be chosen so that the bosonic gluing functions $f^i_{\alpha\beta}$ are functions of the $z$'s only.  It is holomorphically
split if the local coordinates can be chosen so that in addition the fermionic gluing functions $\psi^s_{\alpha\beta}$
are linear in the $\theta$'s.  It is rare to encounter in practice a complex supermanifold that is holomorphically
projected but not holomorphically split, though it is not difficult to construct an example.

As we will see shortly in an example, a generic complex supermanifold is not holomorphically projected. The significance of the question is
that when  holomorphic
projections exist, they tend to be unique or nearly so and may be natural and useful.

A case in point is the moduli space of super Riemann surfaces, which we will call $\MM$.
Explicit computations that have been done to date in superstring 
theory -- including the very beautiful two-loop computations of \cite{DPhmore} -- make
use of the fact that the moduli space $\MM$ admits a holomorphic 
splitting for low genus.  However,  $\MM$ is not holomorphically split in general \cite{DW}.  
In fact, one goal  in \cite{reconsidered}
will be  to make it clear that the existence of a systematic algorithm for 
superstring perturbation theory does not depend on the existence of a holomorphic
splitting.  

For a simple example of a complex supermanifold that cannot be 
holomorphically projected  (see section 4.2.10 of \cite{Manin} for another description 
of this example), let $M$ be the hypersurface
\begin{equation}\label{hypersurface}z_1^2+z_2^2+z_3^2 +\theta_1\theta_2=0\end{equation}
in the projective space $\CP^{2|2}$. The reduced space $M_\red$ is a 
hypersurface in an ordinary projective space $\CP^2$. We write
$\hat z_1\dots \hat z_3$  for homogeneous coordinates of this $\CP^2$ and 
define $M_\red$ as the hypersurface $\h z_1^2+\h z_2^2+\h z _3^2=0$.
A holomorphic projection $\pi:M\to M_\red$ would express the $\hat z_i$ as 
holomorphic functions of the $z_i$ and $\theta_j$.  Since we want $\pi\circ i=1$ where
$i:M_\red\to M$ is the natural inclusion, $\hat z_i$ must coincide with 
$z_i$ at $\theta_1=\theta_2=0$.  So we must have 
$\hat z_i=z_i+\theta_1\theta_2 u_i$ for some $u_i$. The
$u_i$ cannot exist if they are supposed to be holomorphic;  
they would have to  be homogeneous of degree $-1$ under scaling of $z_1\dots|\dots\theta_2$.

The structure theorem of smooth supermanifolds tells us that if $M$ were a smooth
supermanifold, then the $u_i$ would exist.  Here we run into a subtlety that will be
the subject of section \ref{bicomplex}.   It is possible to endow $M$ with a smooth structure --
or more precisely to define a smooth supermanifold $M_{\cs}$ that admits a complex structure
in which it is isomorphic to $M$ -- but there is no truly canonical way to do this.
Postponing this somewhat knotty story to section \ref{bicomplex}, here we will
just define an appropriate $M_\cs$ in our example and show that with this choice, the $u_i$
do exist.  We define $M_\cs$ as a cs supermanifold by starting with coordinates
$t_1\dots t_6|\theta_1\theta_2$, and defining $z_k=t_k+\sqrt{-1}\,t_{k+3},$ 
$\t z_k=t_k-\sqrt{-1} \,t_{k+3}$,
$k=1,\dots,3$.  Then we impose equation (\ref{hypersurface}) along with 
\begin{equation}\label{ombolo} \sum_{k=1}^3 \t z_kz_k=1,\end{equation}
and we divide by the equivalence
\begin{align}\label{zunk} z_k& \to e^{i\alpha}z_k,~~k=1\dots 3 \cr
                                       \theta_s& \to e^{i\alpha}\theta_s~~~s=1,2 \cr
                                        \t z_k&\to e^{-i\alpha}\t z_k,~~k=1\dots 3,\end{align}
  where $\alpha$ is a real parameter (more precisely an even parameter that is real
  modulo the odd variables) and $i=\sqrt{-1}$.   
  If one prefers, one can express all this
  in terms of the $t$'s and $\theta$'s without introducing the $z$'s and $\t z$'s.  This procedure defines a smooth supermanifold
  $M_\cs$ of dimension $4|2$ that admits a complex structure in which it is isomorphic
  to $M$.  The structure theorem of smooth supermanifolds tells us that $M_\cs$ must
  split, and indeed a splitting is given by
   \begin{equation}\label{obolfo}\hat z_i=
   z_i+\theta_1\theta_2 \frac{\t z_i}{2},\end{equation}
   since this condition along with (\ref{hypersurface}) and (\ref{ombolo}) implies that
   $\sum_{k=1}^3\h z_k^2=0$.
      
If $M$ is a complex supermanifold of odd dimension 1, and with no 
odd moduli on which the gluing functions depend, then $M$ is inevitably 
split.  The reason for this is simply
that if a single odd coordinate $\theta$ is the only odd variable that appears 
in the gluing functions of eqn. (\ref{borro}), then inevitably those gluing functions have the split
form ($f$ is independent of $\theta$ since there is no way to make a fermion 
bilinear, and similarly $\psi$ is homogeneous and linear in $\theta$).  
An important example of this in superstring theory is the one-loop dilaton tadpole.
This involves a moduli space of odd dimension 1, so it is naturally 
split and some of the subtleties of superstring perturbation theory do not arise.

\section{Integration On Supermanifolds}\label{superintegration}

We will give two different explanations of what sort of object can be integrated on a supermanifold.  The first explanation is possibly
slightly abstract, but is directly related to the way that a measure on supermoduli space has been extracted in the literature from superconformal
field theory \cite{MNP,EHVerl,superoperator,DPh}.  The second explanation is possibly more concrete and gives a convenient framework for  the supermanifold
version of Stokes's theorem \cite{BL}.  We also will give several descriptions of how to construct objects than can be integrated over suitable submanifolds
of a supermanifold.

\subsection{Sections Of The Berezinian}\label{secber}

The basic idea of the Berezin integral is presumably familiar.  On $\R^{p|*q}$, with bosonic and fermionic coordinates $t^1\dots|\dots\theta^q$, a general
function $g$ can be expanded as a polynomial in the $\theta$'s:
\begin{equation}\label{gork}g(t^1\dots|\dots\theta^q)=g_0(t^1\dots t^p)+\dots+\theta^q\theta^{q-1}\dots\theta^1 g_q(t^1\dots t^p).\end{equation}
We have written explicitly the first and last terms in the expansion.
If $g$ is compactly supported, or at least if  $g_q$  vanishes fast enough at infinity, then the integral of $g$ over $\R^{p|*q}$ is defined as 
\begin{equation}\label{helmbo}\int_{\R^{p|q}}\[\d t^1\dots |\dots\d\theta^q\]  \,g(t^1\dots|\dots\theta^q)=\int_{\R^p}\d t^1\dots\d t^p \,g_q(t^1\dots t^p).
\end{equation}
The $\theta$'s are treated in a purely algebraic fashion, so the question of whether they admit a real structure is immaterial.

To generalize this Berezin integral to a general supermanifold, we want to know what sort of object
 is the ``integration form'' $\[\d t^1\dots|\dots \d\theta^q\]$.  On an ordinary oriented manifold, this would be a differential
form of top degree, but on a supermanifold that is the wrong interpretation.  For example, the formula (\ref{gork}) implies
that if we rescale one of the $\theta$'s by a constant $\lambda$, the symbol $\[\d t^1\dots|\dots\d\theta^q\]$ is multiplied by $\lambda^{-1}$,
rather than by $\lambda$ as one would expect for a differential form.    

To elucidate the meaning of the integration form, we  begin with an approach that is slightly abstract but
actually closely related to formulas in the superstring literature of the 1980's.
First we practice with a vector space.\footnote{The matrix elements of the blocks $B$ and $C$ in eqn. (\ref{defber}) below are odd,
and to allow them to be nonzero, we should work not over a field but over a $\Z_2$-graded ring that has odd elements.}
Let $V$ be a super vector space of bosonic and fermionic dimensions $p|q$.
A basis of $V$ therefore consists of $p$ even vectors $e_1\dots e_p$ and $q$ odd vectors $\rho_1\dots\rho_q$, with the whole collection being
linearly independent.   We abbreviate the basis as $(e_1\dots|\dots\rho_q)$.

The Berezinian of $V$, denoted $\Ber(V)$, is a one-dimensional vector space that one can think of as the space of densities on $V$.  It is defined as follows.
For every basis $(e_1\dots|\dots\rho_q)$ of $V$, there is a corresponding vector in $\Ber(V)$ that we denote as  $\[e_1\dots|\dots\rho_q\]$.
If $(e_1',\dots|\dots,\rho_q')$ is a second basis, related to the first by a linear transformation
\begin{equation}\label{odexo}\begin{pmatrix}e_1'\cr e_2'\cr \vdots \cr - \cr \vdots \cr \rho'_q\end{pmatrix} =
W \begin{pmatrix}e_1\cr e_2\cr \vdots \cr - \cr \vdots \cr \rho_q\end{pmatrix},\end{equation}
then the corresponding two elements of $\Ber(V)$ are related by
\begin{equation}\label{dexo}\[e_1'\dots|\dots,\rho_q'\]=\Ber(W)\,\[e_1,\dots|\dots,\rho_q\].\end{equation}
Here $\Ber(W)$, known as the Berezinian of $W$, is the superanalog of the determinant, in the sense that it possesses the
same multiplicative property: if $W=W_1W_2$, then $\Ber(W)=\Ber(W_1)\Ber(W_2)$.  However, its definition is more subtle than that of the ordinary
determinant. Picking a decomposition of $V$ as $V=V_{\mathrm{even}}\oplus V_{\mathrm{odd}}$, where the summands are of dimension $p|0$ and $0|q$,
we can write a linear transformation of $V$ in block form as 
\begin{equation}\label{defber} W=\begin{pmatrix} A & B\cr C & D\end{pmatrix},\end{equation}
where here $A$ and $D$ are respectively $p\times p$ and $q\times q$ even blocks, while $B$ and $C$ are odd.  The Berezinian
is defined for matrices $W$  such that $D$ is invertible, a condition that is automatically obeyed for the change of basis matrix in (\ref{odexo}).
An explicit formula is
\begin{equation}\label{efber}\Ber(W)=\det(A-BD^{-1}C)\,{\det}^{-1}(D).\end{equation}
If $W$ is upper or lower triangular -- so that $B$ or $C$ vanishes -- then simply
\begin{equation}\label{bact}\Ber(W)=\frac{\det A}{\det\,D}. \end{equation}  It is rather tricky to show that the formula (\ref{efber}) does not depend on the 
chosen decomposition and implies the multiplicative property; for instance, see \cite{Leites}, pp. 15-18; \cite{DeWitt}, section 1.6; or \cite{DEF}, pp. 59-60.
However, it is straightforward to show that the multiplicative property together with (\ref{bact}) implies (\ref{efber}).  This simply follows from the factorization
\begin{equation}\label{formfact}\begin{pmatrix}A&B\cr C&D\end{pmatrix}=\begin{pmatrix}A-BD^{-1}C ~~& B\cr 0 & D\end{pmatrix}\begin{pmatrix}1& 0 \cr D^{-1}C~~ & 1\end
{pmatrix}.\end{equation}
Similarly, the factorization
\begin{equation}\label{tormfact}\begin{pmatrix}A&B\cr C&D\end{pmatrix}=\begin{pmatrix}A~~& 0\cr C~~ & D-CA^{-1}B\end{pmatrix}\begin{pmatrix}1& A^{-1}B \cr 0~~ & 1\end{pmatrix}\end{equation}
implies another formula $\Ber(W)=\det A\cdot {\det}^{-1}(D-CA^{-1}B)$.

Now let $M$ be a compact supermanifold\footnote{To avoid having to include in the formulas some minus signs  which could be confusing on first
reading, we will assume that the reduced space $M_\red$ of $M$ is oriented.    When we identify $U_\alpha\subset M$ with an open
subset of $\R^{p|* q}$, the orientation of $M_\red$ determines an orientation of the reduced space of $\R^{p|*q}$. 
This lets us view the quantity $\d t^1\dots \d t^p$ in
(\ref{helmbo}) and related formulas as a differential form rather than a density.}  of dimension $p|q$, as described in section 
\ref{smooth}.  Let $T^*M$ be the cotangent bundle of $M$.  We will introduce on $M$ a line bundle known
as its Berezinian line bundle $\Ber(M)$.    $\Ber(M)$ is defined by saying that every local coordinate
system $T=t^1\dots|\dots\theta^q$ on $M$ determines a local trivialization of $\Ber(M)$ that we denote $\[\d t^1\dots|\dots\d\theta^q\]$.
Moreover, if $\t T=\t t^1 \dots|\dots \t \theta^q$ is another coordinate system, then the two trivializations of $\Ber(M)$ are related by
\begin{equation}\label{hockey}\[\d  t^1\dots|\dots\d\theta^q\]=\Ber\left(\frac{\partial  T}{ \partial \t T}\right) \[\,\d\t t^1\dots|\dots\d\t \theta^q\].\end{equation}
Here $\partial T/\partial\t T$ is the matrix of derivatives of $( t^1\dots|\dots \theta^q) $ with respect to $(\t t^1\dots|\dots\t\theta^q)$.

We claim that what can be naturally integrated over $M$ is a section of $\Ber(M)$.  To show this, first let $\sigma$ be a section of $\Ber(M)$
whose support is contained in a small open set $U$ on which we are given local coordinates $t^1\dots|\dots\theta^q$, establishing
an isomorphism of $U$ with an open set in $\R^{p|*q}$.  This being so,  we can view $\sigma$ as a section of the Berezinian of $\R^{p|*q}$. This
Berezinian is trivialized by the section $[\d t^1\dots|\dots \d\theta^q]$ and $\sigma$ must be the product of this times some function $g$:
\begin{equation}\label{odzo}\sigma=[\d t^1\dots|\dots\d \theta^q\]\,g(t^1\dots|\dots\theta^q).\end{equation}
So we define the integral of $\sigma$ to equal the integral of the right hand side of eqn. (\ref{odzo}):\begin{equation}\label{helmox}\int_M\sigma=\int_{\R^{p|*q}}\[\d t^1\dots|\dots\d\theta^q\]\,g(t^1\dots|\dots\theta^q).\end{equation}
The integral on the right is the naive Berezin integral (\ref{helmbo}).
For this definition to make sense, we need to check that the result does not depend on the coordinate system $t^1\dots|\dots\theta^q$
on $\R^{p|*q}$ 
that was used in the computation.  This follows from the rule (\ref{hockey}) for how the symbol $\[\d t^1\dots|\dots\d\theta^q\]$ transforms
under a change of coordinates.  The Berezinian in this formula is analogous to the usual Jacobian in the transformation law of an ordinary integral under
a change of coordinates.  For more detail, see for instance \cite{DeWitt}, pp. 40-1; \cite{Rogers}, Theorem 11.2.3; or \cite{DEF}, p. 80.  

So far, we have defined the integral of a section of $\Ber(M)$ whose support is in a sufficiently small region in $M$.  To reduce the general case
to this, we pick a cover of $M$ by small open sets $U_\alpha$, each of which is isomorphic to an open set in $\R^{p|*q}$, and
we  use the existence of a partition of unity.  Just as on a bosonic manifold, one can find bosonic functions $h_\alpha$
on $M$ such that each $h_\alpha$ is supported in the interior of $U_\alpha$ and $\sum_\alpha h_\alpha=1$.   Then we write $\sigma=\sum_\alpha
\sigma_\alpha$ where $\sigma_\alpha =\sigma h_\alpha$.  Each $\sigma_\alpha$ is supported in $U_\alpha$, so its integral can be defined
as in (\ref{helmox}).  Then we define $\int_M\sigma=\sum_\alpha\int_M\sigma_\alpha$.  That this does not depend on the choice of the open cover
or the partition
of unity follows from the same sort of arguments used in defining the integral of a differential form on an ordinary manifold.  For example,
see Theorem 11.3.2 of \cite{Rogers}.

\subsubsection{Relevance To Superstring Perturbation Theory}\label{relevance}

All this is perhaps a little abstract and in a sense a tautology: we simply postulated the right transformation law for the symbol
$\[\d t^1\dots|\dots\d \theta^q\]$ so that the integral does not depend on a choice of coordinates.  One reason that the 
definition is useful is that it matches the superstring literature, a fact that will be relevant in \cite{reconsidered}.  In applications to superstring 
perturbation theory, it is  helpful
to know the following characterization of $\Ber(M)$.  To every basis $\partial_{t^1}\dots |\dots\partial_{\theta^q}$ of the tangent space
of $M$, a section $\sigma$ of $\Ber(M)$ assigns a number, which we will denote as $\sigma(\partial_{t_1}\dots|\dots\partial_{\theta^q})$.
We define this mapping by saying that it is linear in $\sigma$ and for $\sigma = [\d t^1\dots|\dots\d \theta^q\]$, we have
$\sigma(\partial_{t^1}\dots|\dots\partial_{\theta^q})=1$.  Under a change of basis, we require
\begin{equation}\label{zelb}\sigma(\partial_{\t t^1}\dots|\dots\partial_{\t \theta^q})= \sigma(\partial_{t^1}\dots|\dots\partial_{\theta^q})\,\,
\Ber\left(\frac{\partial T}{\partial\t  T}\right),\end{equation}
which is compatible in a natural way with  (\ref{hockey}).  Conversely, a function on bases of the tangent space that transforms in 
this way is equivalent to a section of $\Ber(M)$.  In 
perturbative string theory, $M$ will be supermoduli space, the basis $\partial_{t^1}\dots|\dots\partial_{\theta^q}$
will correspond to a basis of super Beltrami differentials, and a section $\sigma$ of $\Ber(M)$  will
be defined by saying that $\sigma(\partial_{t^1}\dots|\dots\partial_{\theta^q}) $ is the value of
the worldsheet path integral with superghost insertions corresponding to those super Beltrami differentials.  

The construction in this section shows that a smooth section of $\Ber(M)$ can always be integrated over a compact supermanifold $M$.
So, in the context of superstring perturbation theory, given that superconformal field theory can be used to define a section of $\Ber(M)$,
the only possible difficulty comes from the lack of compactness of supermoduli space, arising
from the infrared region. 

\subsubsection{A Note On Notation}\label{notenot}

A point to stress is that the section of the Berezinian that we have written as $[\d t^1\dots |\dots\d\theta^q]$ is an irreducible object; we have not built it by multiplying differential forms $\d t$ and $\d \theta$.  We have not yet even introduced differential forms.  The notation is meant to evoke the idea of a volume form, but
the bracket  $[~~~~]$ surrounding the $\d t$'s and $\d \theta$'s is a warning that the symbols inside the bracket have only
an abstract meaning.

In section \ref{informs}, we will introduce differential (and integral) forms on a supermanifold; when we do so, for an odd variable $\theta$, the one-form $\d\theta$
will be an even variable.  So for example we will have $\d\theta\d\theta'=d\theta'\d\theta$ and  $(\d\theta)^2\not=0$.

By contrast, the symbol $[\d t^1\dots |\dots \d\theta^q]$ is odd under exchange of any two $\theta$'s. This assertion is a special case of (\ref{dexo}) or (\ref{hockey}).
The exchange of two $\theta$'s is a coordinate transformation with $\Ber(\partial \tilde T/\partial T)=-1$, as one can see from (\ref{bact}).  

\subsubsection{Integrating Over The Fibers Of A Fibration}\label{fibration}

Consider a space $\R^{p+p'|*(q+q')}$, which we decompose in some fashion as $\R^{p|*q}\times \R^{p'|*q'}$.  
The Berezin integral of a smooth, compactly supported function on $\R^{p+p'|*(q+q')}$
can be performed by first integrating over $\R^{p|*q}$ to get a function on $\R^{p'|*q'}$, which one then integrates over
$\R^{p'|*q'}$.    In fact, if we integrate first over the odd variables, then the statement just made reduces to the fact
that a similar ordinary integral on $\R^{p+p'}=\R^p\times \R^{p'}$ can be performed by 
integrating first over $\R^p$
and then over $\R^{p'}$.

Now consider a general fibration of supermanifolds $\pi:X\to N$, with fiber $M$.  We claim that 
by integrating over the fibers of $\pi$, one can define a natural linear map $\sigma\to \pi_*(\sigma)$
from a section $\sigma$ of $\Ber(X)$ to a section $\pi_*(\sigma)$ of $\Ber(N)$, obeying the fundamental relation
\begin{equation}\label{edzo}\int_X\sigma=\int_N\pi_*(\sigma).\end{equation}
Just as in the original definition of integration on a supermanifold,  by linearity of the integral it suffices to consider the case
that the support of $\sigma$ is in a sufficiently small open set in $X$.  In particular, we can assume that the support of $\sigma$ projects
in $N$ to an open set that is isomorphic to an open set in some $\R^{p|*q}$.  We can also assume that the fibration $\pi$ is
a product when restricted to this open set, and that along the fibers of $\pi$, the support of $\sigma$ is contained in an open
set that is isomorphic to an open set in some $\R^{p'|*q'}$.  So the definition of the operation $\pi_*$ and the verification of (\ref{edzo})
reduce to the special case mentioned in the last paragraph. 

 \subsection{Differential And Integral Forms}\label{informs}

Now we are going to describe a phenomenon that in a sense is at the root of the picture-changing phenomenon in superstring
perturbation theory.

\subsubsection{Clifford Algebras}\label{clifford}

Let  $V\cong \R^{0|p}$ be a purely odd vector space of dimension $p$.  Let $\zeta^1,\dots,\zeta^p$ be a basis of $V$
and let $\eta_1,\dots,\eta_p$ be a basis of the dual space $V^*$.  There is a natural nondegenerate symmetric bilinear form on the direct sum $V\oplus V^*$ which we can  write
\begin{equation}\label{canwrite}(\zeta^i,\zeta^j)=(\eta_i,\eta_j)=0,~~(\zeta^i,\eta_j)=(\eta_j,\zeta^i)=\delta^i_j.\end{equation}
This quadratic form does not depend on the chosen basis.

Given this nondegenerate quadratic form on $V\oplus V^*$, we can ``quantize'' by introducing the corresponding Clifford algebra.  This simply means that we promote
the $\zeta^i$ and $\eta_j$ to operators that are supposed to obey the anticommutation relations
\begin{equation}\label{orzok} \{\zeta^i,\zeta^j\}=0=\{\eta_i,\eta_j\},~~\{\zeta^i,\eta_j\}=\delta^i_j.\end{equation}
Anticommutation (rather than commutation) relations are natural since  $\zeta^i$ and $\eta_j$ are odd variables. 
An irreducible module\footnote{A module for the Clifford algebra is simply a vector space on
which the algebra acts, analogous to  a representation of a group.}
 $\S$ for this Clifford algebra (or any Clifford algebra of even rank) is unique up to isomorphism.  We can construct
it by starting with a vector $|\neg\downarrow\rangle$ annihilated by the $\eta_i$; then the states $\zeta^{i_1}\dots \zeta^{i_k}|\neg\downarrow\rangle$, $i_1<\dots<i_k,
\,k=0,\dots,p$ furnish a basis of $\S$.  Alternatively, we can start with a state $|\neg\uparrow\rangle$ that is annihilated by the $\zeta^j$, and build
a basis by acting on $|\neg\uparrow\rangle$ with the $\eta_i$.   The two constructions are equivalent, since $|\neg\uparrow\rangle= \zeta^1\zeta^2\dots \zeta^p|\neg\downarrow\rangle$
can be reached from $|\neg\downarrow\rangle$ after finitely many steps, and vice-versa.

We would like to interpret this construction more geometrically, but in doing so we may as well consider a more general situation involving a purely bosonic manifold 
 $M$  of dimension $p$.  Roughly speaking, we want to consider functions on the tangent bundle $TM$ of $M$.  But there is a very important
twist: we want to consider the fiber directions of the tangent bundle to be fermionic rather than bosonic.    The tangent bundle with this twist is frequently denoted as ${\Pi} TM$,
where the symbol ${\Pi}$ stands for reversal of statistics in the fiber directions; in the literature, this is often called reversal of parity.  If $t^1\dots t^p$ are local coordinates on $M$,
then to give a local coordinate system on ${\Pi} TM$, we need to double the coordinates, adding a second set $\d t^1\dots\d t^p$.  These now are fermionic variables since we 
have taken the fiber coordinates of ${\Pi} TM$ to be odd.  A general function on ${\Pi} TM$ has an expansion in powers of the $\d t^i$, and of course this is a finite expansion
since these variables anticommute.  A term of order $k$
\begin{equation}\label{ohomo} \sum_{i_1\dots i_k}a_{i_1\dots i_k}(t^1\dots t^p)\d t^{i_1}\dots\d t^{i_k} \end{equation}
corresponds to what usually is called a $k$-form on $M$.  So taking all values of $k$ from 0 to $p$, the functions on ${\Pi} TM$ correspond to the whole space of differential
forms on $M$, often denoted $\Omega^*(M)$.

At any point $m\in M$, we can easily find a Clifford algebra acting on the functions on the fiber of ${\Pi} TM$ at $m$.
We take $\zeta^i$ to be  multiplication by $\d t^i$,
\begin{equation}\label{curtsy}\psi \to \d t^i\wedge \psi\end{equation}
(we usually omit the wedge product symbol) and
we take $\eta_j$ to be the corresponding derivative operator
\begin{equation}\label{styro}\eta_j=\frac{\partial}{\partial \d t^j}.\end{equation}
In differential geometry, this operator is usually called
the operator of contraction with $\partial_{t^j}$,
\begin{equation} \label{urtsy}\omega\to \i_{{\tiny\partial}_{t^j}}\omega,\end{equation}
but it is simpler to think of it as the derivative with respect to the bosonic variable $\d t^j$.

The $\zeta^i$ and $\eta_j$ as just defined obey the fiberwise Clifford algebra $\{\zeta^i,\eta_j\}=\delta^i_j$, with other anticommutators vanishing.
Differential forms on $M$ or equivalently functions on ${\Pi} TM$ give a natural module for this family of Clifford algebras; any other irreducible module would be
equivalent (except that globally  one could consider differential forms on $M$ with values in a line bundle).

The exterior derivative operator corresponds to a simple odd vector field on ${\Pi} TM$:
\begin{equation}\label{zomb}\d=\sum_i \zeta^i\frac{\partial}{\partial t^i} =\sum_i \d t^i\frac{\partial}{\partial t^i}.\end{equation}
It obeys $\d^2=0$.  Since it has degree 1 under scaling of the fiber coordinates of ${\Pi} TM$,  
it maps forms of degree $k$ -- that is functions homogeneous of degree $k$ in the odd variables $\d t^i$ -- to forms of degree $k+1$.  When a differential form is interpreted as a function $\omega(t^1\dots t^p|\d t^1\dots \d t^p)$,
the wedge product of forms becomes obvious -- it is simply the multiplication of functions.  That is why we usually omit the wedge product symbol.
The wedge product and exterior derivative are related by
\begin{equation}\label{udofl} \d(\omega\cdot \nu)= \d\omega \cdot \nu + (-1)^{|\omega|}\omega\cdot \d\nu,\end{equation}
where $|\omega|=0$ or 1 for $\omega$ even or odd.

\subsubsection{Weyl Algebras}\label{weyl}

Now let us repeat this exercise for a purely even vector space $W\cong \R^{q|0}$.  Again we pick a basis $\alpha^1,\dots,\alpha^q$
of $W$ and a dual basis $\beta_1,\dots,\beta_q$ of the dual space $W^*$.  Because the $\alpha$'s and $\beta$'s are now even,  it is
more useful to introduce a skew-symmetric rather than symmetric bilinear form on 
$W\oplus W^*$:
\begin{equation}\label{hokey} \l \alpha^i,\alpha^j\r = \l \beta_i,\beta_j\r = 0,~~ \l \beta_j,\alpha^i\r= -\l \alpha^i,\beta_j\r  = \delta^i_j. \end{equation}
Upon quantization, the even variables $\alpha^i$ and $\beta_j$ obey a Weyl algebra rather than a Clifford algebra:
\begin{equation}\label{dokey}[\alpha^i,\alpha^j]=[\beta_i,\beta_j]=0,~~[\beta_j,\alpha^i]=\delta^i_j. \end{equation}
In contrast to the finite-dimensional Clifford algebra, the finite-dimensional Weyl algebra has many irreducible modules.
We can postulate a state $|\neg\downarrow\rangle$ that is annihilated by the $\beta$'s.  Then a module $\V$ for the Weyl algebra can be constructed
by acting repeatedly with $\alpha$'s.  A basis of this module consists of states of the form
\begin{equation}\label{ubasis}\alpha^{i_1}\alpha^{i_2}\dots\alpha^{i_k}|\neg\downarrow\rangle,~~k\geq 0. \end{equation}
Because the $\alpha$'s are commuting variables, there is no upper bound on $k$.  Alternatively, we can postulate the existence of
a state $|\neg\uparrow\rangle$ annihilated by the $\alpha$'s, in which case we form a module $\V'$ with a basis of states 
\begin{equation}\label{nubasis} \beta_{j_1}\beta_{j_2}\dots\beta_{j_k}|\neg\uparrow\rangle,  ~~k\geq 0. \end{equation}
The two modules are inequivalent, as $\V$ contains no state annihilated by the $\alpha$'s and $\V'$ contains no state annihilated by the $\beta$'s.
Of course one can form a mixture of the two cases (and we will discuss such mixtures in section \ref{irred}).  But these two
cases are of particular importance.  

It is convenient to construct both modules by representing the $\alpha^i$ as multiplication operators and the $\beta_j$ as derivatives:
\begin{equation}\label{odolfo} \beta_j=\frac{\partial}{\partial \alpha^j}. \end{equation}
If we do this, then the two modules differ by the classes of functions allowed.  To obtain $\V$, we consider polynomial functions of the
$\alpha$'s, with the state $|\neg\downarrow\rangle$ corresponding to the function $f(\alpha^1\dots\alpha^q)=1$, which is annihilated by the $\beta$'s.
To construct $\V'$, we need a state $|\neg\uparrow\rangle$ that is annihilated by the operation of multiplication by $\alpha^i$, $i=1\dots q$.
As a function of the $\alpha$'s, $|\neg\uparrow\rangle$ corresponds to a delta function supported at $\alpha^1=\dots=\alpha^q=0$.  Certainly
the state $\delta^q(\alpha^1\dots\alpha^q)$ is annihilated by multiplication by any of the $\alpha$'s.  Acting repeatedly with the $\beta$'s, we
see that $\V'$ is spanned by distributions supported at the origin; a basis of $\V'$  consists of states of the form
\begin{equation}\label{nudolf}\frac{\partial}{\partial \alpha^{i_1}}\frac{\partial}{\partial \alpha^{i_2}}\dots\frac{\partial}{\partial \alpha^{i_k}}\delta^q(\alpha^1
\dots\alpha^q),~~k\geq 0.\end{equation}

To interpret this more geometrically, we introduce a purely fermionic supermanifold $M$ of dimension $0|q$.  We may as well take $M$ to
be $\R^{0|q}$, with coordinates $\theta^1\dots\theta^q$.  We introduce an ``exterior derivative'' $\d$,  which we consider to be odd, just as in the bosonic case, so that it obeys, for example, \begin{equation}\label{holes}
\d(\theta^1\theta^2)=\d\theta^1\cdot \theta^2
-\theta^1\cdot \d\theta^2.\end{equation}  So the objects $\alpha^i=\d \theta^i$, which we will call one-forms, are even variables.
  The $\beta_j$ then become 
contraction operators
\begin{equation}\label{thelc}\beta_j=\i_{{\tiny\partial}_{\theta^j}}=\frac{\partial}{\partial\alpha^j}.\end{equation}

Just as in the bosonic case, a differential form on $\R^{0|*q}$ can be interpreted as a function $\omega(\d\theta^1\dots\d\theta^q|\theta^1\dots
\theta^q)$ (our convention is to list the even  variables first), the only subtlety being that the class of functions considered depends on
whether we want the module $\V$ or $\V'$.  For $\V$, the functions have polynomial dependence on the $\d\theta^i$, but for $\V'$,
they are delta functions supported at $\d\theta^i=0$.  Also as in the bosonic case, the wedge product of forms
represented by two functions $\omega$ and $\nu$ is simply represented by the product $\omega\nu$ of the two functions.
Here, however, there is a subtlety: one can multiply two polynomials, and one can multiply a distribution by a polynomial, but one
cannot multiply two distributions.  So the wedge product makes sense as a map $\V\times \V\to\V$, and also as a map $\V\times \V'\to\V'$,
but there is no way to multiply two elements of $\V'$.

On either module $\V$ or $\V'$, we can define an exterior derivative operator
\begin{equation}\label{extder}\d=\sum_i \alpha^i\frac{\partial}{\partial\theta^i}=\sum_i\d\theta^i\frac{\partial}{\partial\theta^i}.\end{equation}
Wherever the wedge product is defined, the wedge product and the exterior derivative obey the relation (\ref{udofl}).

\subsubsection{Forms On Supermanifolds}\label{genc}

We have treated separately the  purely bosonic and purely fermionic cases, but there is no problem to consider in the same way
a general supermanifold $M$, say of dimension $p|q$.  Given local coordinates $t^1\dots|\dots\theta^q$, we introduce
the correponding one-forms  $\d t^1\dots \d t^p$ and $\d\theta^1\dots\d\theta^q$, which are respectively fermionic and bosonic.  Forms will
correspond  to functions $\omega(t^1\dots\d\theta^q|\theta^1\dots\d t^p)$ of all the variables, including the $\d t$'s and $\d\theta$'s,  or in other words
to functions on ${\Pi} TM$.  The only subtlety is what sort of functions are allowed.  We get what one may call  differential forms on $M$ if
we require $\omega$ to have polynomial dependence on the even one-forms $\d\theta^i$.  And we get integral forms (a concept that originated in \cite{BL})  if
we require that in its dependence on $\d\theta^1\dots \d\theta^q$, $\omega$ is a distribution supported at the origin.  We write
$\Omega^*(M)$ for the differential forms and $\Omega^*_{\mathrm{int}}(M)$ for the integral forms.  
We take diffeomorphisms to act on the variables $\d t^i$ and $\d\theta^j$ by the usual chain rule.  For example, if $\t t^1\dots|\dots\t\theta^q$
is another coordinate system, then 
\begin{align}\label{onx}\d t^i =&\sum_k \frac{\partial t^i}{\partial\t t^k}\d\t t^k+\sum_s \frac{\partial t^i}{\partial \t\theta^s}\d \t\theta^s\cr
    \d \theta^r =&\sum_k \frac{\partial\theta^r}{\partial\t t^k}\d\t t^k+\sum_s \frac{\partial \theta^r}{\partial \t\theta^s}\d \t \theta^s. \end{align}
Forms with only polynomial dependence on $\d\theta^1\dots\d\theta^q$ are mapped to forms of the same type by a change of coordinates;
the same is true for forms with support    
only at $\d\theta^1=\dots=\d\theta^q=0$. Hence the space of
 differential forms and the space of integral forms are each invariant under gluing.  So 
 these spaces are globally-defined on any supermanifold $M$, even though our initial definition used a local coordinate system.

A wedge product of two differential forms
or of a differential form and an integral form is defined by multiplying the corresponding functions.  
The exterior derivative is defined in the obvious way as a vector field on ${\Pi} TM$
\begin{equation}\label{ormoc}\d=\sum_i \d t^i\frac{\partial}{\partial t^i}+\sum_j\d\theta^j\frac{\partial}{\partial\theta^j}\end{equation}
that obeys $\d^2=0$ and satisifies the usual relation (\ref{udofl}).

Differential forms are graded by degree in the usual way -- a function $f(t^1\dots|\dots\theta^q)$ on $M$ is a zero-form,
and multiplying $k$ times by one-forms $\d t^i$ or $\d\theta^j$  gives a $k$-form.  In the space of differential forms, there are 
forms of lowest degree -- namely degree 0 -- but there are no forms of highest degree.  If we introduce a scaling symmetry of ${\Pi} TM$
that acts trivially on $t^1\dots|\dots\theta^p$ but scales the fiber coordinates by a common factor $\lambda$
\begin{equation}\label{dromo}\d t^1\dots|\dots\d\theta^q\to \lambda\d  t^1\dots|\dots\lambda \d\theta^q,\end{equation} then a $k$-form
scales as $\lambda^k$.

In the case of integral forms, we will say that for any function $f$ on $M$, the object
\begin{equation}\label{topf}f(t^1\dots|\dots\theta^q)\d t^1\dots \d t^p\delta^q(\d\theta^1\dots\d\theta^q)\end{equation}
 is a top form.  Such a form is annihilated by multiplication by $\d t^i$ or $\d\theta^j$. 
A form obtained by acting $k$ times with operators $\partial/\partial(\d t^i)$ or $\partial/\partial (\d\theta^j)$ will be called a form of
codimension $k$.   Again there is no upper bound on $k$, so in the space of integral forms, there are top forms but no bottom forms.
Under the scaling symmetry of ${\Pi} TM$, a top form scales as $\lambda^{p-q}$, and a form of codimension $k$ scales as $\lambda^{p-q-k}$.

Differential forms can be multiplied, and one can do many other things with them,  but they cannot be integrated, roughly because there is no top form. 
To be more exact, if $\omega$ is a $p$-form on $M$, then by essentially the ordinary definitions of calculus,\footnote{If $N$ is of dimension $p|0$, we can 
parametrize it locally by bosonic variables $s^1\dots s^p$.  In terms of these variables, $\omega$ becomes an ordinary $p$-form on an ordinary  $p$-dimensional manifold
and we integrate it in the usual way.  It does not matter that $s^1\dots s^p$ can only be defined locally; as usual, we write $\omega$ as a sum of $p$-forms $\omega_\alpha$
each of which is supported in a small open set $U_\alpha$ in which suitable coordinates exist.} 
$\omega$ can be integrated over a purely bosonic oriented submanifold $N\subset M$ of the right dimension, in fact a submanifold of dimension $p|0$.  But $\omega$ cannot be integrated over any submanifold with positive odd dimension.

Integral forms can be integrated over $M$, as we will discuss shortly (and more generally over submanifolds of purely bosonic codimension, as we will
see in section \ref{intsub}), but they cannot be multiplied.

The relation between differential forms and integral forms is a prototype for the notion of different ``pictures'' in superstring theory.
The concept of different pictures has roots \cite{nst,bors}  in the early days of what developed into superstring theory, and was interpreted in
\cite{FMS} in terms of the existence of inequivalent modules for the Weyl algebra.

\subsection{Integration Of Integral Forms}\label{integintegral}

We will give two related explanations of how integral forms of top degree can be integrated.  

\subsubsection{Integration On ${\Pi} TM$}\label{pint}

One approach starts with a basic difference between $M$ and ${\Pi} TM$.   On $M$, there is  in general no natural
way to pick a section of the Berezinian, but on ${\Pi} TM$ there is always a natural choice because of the way the
variables come in bose-fermi pairs.  For every $t$, there is a $\d t$, and for every $\theta$, there is a $\d \theta$, in each case with
opposite statistics.  Thinking of the whole collection $t^1\dots\d\theta^q|\theta^1\dots\d t^p$ as a local coordinate system on ${\Pi} TM$, the 
corresponding object
\begin{equation}\label{doofus}  [\d t^1\dots \d(\d\theta^q)|\d\theta^1\dots \d(\d t^p)]  \end{equation}
is independent of the underlying choice of coordinates $t^1\dots|\dots\theta^q$ on $M$ and gives a natural section of $\Ber({\Pi} TM)$.   For example, if we rescale one of the even
coordinates of $M$  by $t\to \lambda t$, then one of the odd coordinates on ${\Pi} TM$ is similarly rescaled by $\d t\to \lambda \d t$.
The symbol $ [\d t^1\dots \d(\d\theta^q)|\d\theta^1\dots \d(\d t^p)] $ changes by the Berezinian of the change of coordinates,
according to (\ref{hockey}).  The relevant Berezinian is 1 (essentially because $\det D$ is in the denominator in (\ref{bact}) while $\det A$ is in the
numerator), so
the symbol $ [\d t^1\dots \d(\d\theta^q)|\d\theta^1\dots \d(\d t^p)] $ is invariant under this change of coordinates and indeed it is invariant
under any change of coordinates on $M$.  

To streamline our notation, we will abbreviate the whole set of coordinates $t^1\dots|\dots\theta^q$ on $M$ as $x$, and write
just $\D (x,\d x)$ for  $[\d t^1\dots \d(\d\theta^q)|\d\theta^1\dots \d(\d t^p)] $.  Similarly, we regard an integral form $\omega$ on $M$ as a function $\omega(x,\d x)$
on ${\Pi} TM$.  Now we define the integral of $\omega$ over $M$ as a Berezin integral on ${\Pi} TM$:
\begin{equation}\label{odz} \int_M\omega=\int_{{\Pi} TM}\D (x,\d x)\,\,\omega(x, \d x).\end{equation}

It is crucial here that $\omega$ is an integral form rather than a differential form.
Because $\omega(x)$ has compact support as a function of even variables $\d\theta^1\dots\d\theta^q$ (and in fact is
a distribution with support at the origin),
the integral over those variables makes sense.  A similar approach to integrating a differential form on $M$ would not make
sense, since if $\omega(x)$ is a differential form, it has polynomial dependence on $\d\theta^1\dots\d\theta^q$ and the integral over
those variables does not converge.   (This is why differential forms can only be integrated on purely bosonic submanifolds.)
The formula (\ref{odz}) for integration of $\omega$ makes sense for an integral  form $\omega$ of any
codimension, but if $\omega$ has positive codimension, then this formula vanishes.

Since we have expressed the integral of an integral form in terms of the Berezin integral in a space with twice as many variables, 
the reader may wonder if we are making
any progress.  Why not stick with the original Berezin integral on $M$?  One answer is that in the framework of integral forms, one can
 formulate a supermanifold analog of Stokes's theorem.  A related answer is that the formulation with integral forms turns out
to be useful in superstring perturbation theory.

\subsubsection{Some Technical Remarks}\label{technical}

Two technical remarks  are unfortunately difficult to avoid here.  First, we need to clarify what is meant
by integration over the bosonic variable $\d\theta$.  If we are on a real supermanifold, then $\d\theta$ is an ordinary
real variable, and integration over $\d\theta$ requires no special explanation. However, we are mainly interested in cs
manifolds, so that a fermionic variable $\theta$ has no real structure, and we cannot claim that $\d\theta$ is a real  variable.
This being so, we will have to interpret integration over $\d\theta$ as a formal operation, constrained so that one
can integrate by parts, just like the original Berezin integral over $\theta$.
The only ``functions'' of $\d\theta$ that we need to integrate are distributions supported at the origin.  So we define
the integral over $\d\theta$ by
\begin{equation}\label{zebo}\int[\d (\d\theta)]\frac{\partial^n}{\partial(\d\theta)^n}\delta(\d\theta)=\delta_{n,0}.\end{equation}
Second, we want to interpret the symbol $[\d(\d\theta)]$ as a section of the Berezinian, not a density.
So under a change of variables $\theta\to\lambda\theta$, $\lambda\in \C$, which induces $\d\theta\to \lambda\d\theta$, we want
 $[\d(\d\theta)]\to \lambda[\d(\d\theta)]$, with no absolute value; this is a special case of (\ref{dexo}).  So consistency with (\ref{zebo}) requires
 \begin{equation}\label{oxco}\delta(\lambda\d\theta)=\frac{1}{\lambda}\delta(\d\theta),\end{equation} again
 with no absolute value. Another consequence becomes clear if there are two odd variables $\theta$ and $\theta'$.  We would like
 \begin{equation}\label{zobort}\int [\d(\d\theta)\, \d(\d\theta')] \,\delta(\d\theta)\delta(\d\theta')=1.\end{equation}
 On the other hand, the integration measure $[\d(\d\theta)\,\d(\d\theta')]$ is odd in $\d\theta$ and $\d\theta'$.  So
 the dual delta functions are also anticommuting
 \begin{equation}\label{obort} \delta(\d\theta)\delta(\d\theta')=-\delta(\d\theta')\delta(\d\theta).\end{equation}
   Thus, the calculus of distributional functions of $\d\theta$ and integrals over them is really
 a formal algebraic machinery, like the Berezin integral. 
 
Though this will not be important in the present notes,  
one might wonder how to interpret the  above formulas if $M$ is a real supermanifold, so that we hope to interpret the $\d\theta$'s as real
 variables.  The unfamiliar signs in eqns. (\ref{oxco}) and (\ref{obort}) mean that the symbol $\delta(\d\theta)$ differs slightly from its usual meaning.
 Instead of defining a delta function as a linear function on smooth functions, as is common, we define it as a linear function on smooth differential forms (mathematically,
 an object of this kind is called a current).  To explain the idea,
 let $\R$ be a copy of the real line but with no chosen orientation, and let $\Omega^1(\R)$ be the space of smooth one-forms on $\R$.  An element
 of $\Omega^1(\R)$ is an expression $f(x)\,\d x$ where $f(x)$ is a smooth function.  Such an expression cannot be integrated until we pick an orientation on $\R$.
 We interpret $\delta(x)$ as a linear function on $\Omega^1(\R)$ that maps $f(x)\,\d x$ to $f(0)$. 
 With this interpretation of $\delta(x)$, we have $\delta(-x)=-\delta(x)$, since $\d(-x)=-\d(x)$.  Roughly speaking, $\delta(x)$ in this sense differs
 from the usual $\delta(x)$ by a choice of orientation of the normal bundle to the submanifold $x=0$.
 
 Similarly, with this interpretation, the minus sign in eqn. (\ref{obort}) is natural.  Let $\R$ be a copy of the real line parametrized
 by $x$, and let $Z$ be any manifold.  We define $\delta(x)$ as a map from smooth $k$-forms on $\R\times Z$ to smooth $k-1$-forms
 on $Z$ as follows: $\delta(x)$ annihilates a form that does not contain $\d x$, and it maps $\psi=\d x \wedge \omega$ (for any form $\omega$),
 to the restriction of $\omega$ to $\{0\}\times Z$, where $\{0\}$ is the point $x=0$ in $\R$.  This operation would be integration of 
 $\delta(x)\psi$  over the
 first factor of $\R\times Z$, with a standard interpretation of $\delta(x)$,
 except that integration would require an orientation of that first factor, which we have not assumed.
  If $y$ is another real coordinate, we define
 $\delta(y)$ in the same way.  Now let us consider smooth forms on a product $\R_x\times \R_y$, where $\R_x$ and $\R_y$ are factors of 
 $\R$ parametrized respectively by $x$ and by $y$.  We understand the product $\delta(x)\delta(y)$ to represent successive action of the operator
 $\delta(y)$, mapping smooth $k$-forms on $\R_x\times \R_y$ to smooth $k-1$-forms on $\R_x\times \{y=0\}$, and $\delta(x)$, mapping smooth
 $k-1$-forms on $\R_x$ to smooth $k-2$-forms at the point $x=y=0$.  $\delta(y)\delta(x)$ is understood similarly.  With this meaning
 of the symbols, we have $\delta(x)\delta(y)=-\delta(y)\delta(x)$; indeed, $\delta(y)\delta(x)$ maps the form $\d x\wedge \d y$ to $+1$
 and $\delta(x)\delta(y)$ maps it to $-1$.

\subsubsection{Equivalence With The Berezin Integral on $M$}\label{dirred}

We have defined integration of an integral form $\omega$ on $M$ by means of a Berezin integral over   ${\Pi} TM$
of $\D(x,\d x)\,\omega(x,\d x)$, where $\omega(x,\d x)$ is the function on ${\Pi} TM$ corresponding to the integral form
$\omega$, and $\D(x,\d x)$ is the natural integration measure of ${\Pi} TM$. 

On the other hand, we also have a fibration $\pi:{\Pi} TM\to M$.  And as explained in section \ref{fibration},
a Berezin integral on the total space of a fibration can be performed by first integrating over the fibers of a fibration.  So
if $\omega$ is an integral form, we can define a section $\sigma$ of $\Ber(M)$ by acting with $\pi_*$ on the section
$\D(x,\d x)\,\omega(x,\d x)$ of $\Ber({\Pi} TM)$:
\begin{equation}\label{budz}\sigma=\pi_*\left(\D(x,\d x)\,\omega(x,\d x)\right).\end{equation}
This operation is defined for all integral forms $\omega$, but if $\omega$ has positive codimension then $\sigma=0$.
(For $\omega$ to have positive codimension, it is a linear combination of terms that are  either missing an odd variable $\d t^i$ for some $i $
and vanish
upon integration over $\d t^i$, or are proportional to some $\partial^n\delta(\d \theta^s)$, $n>0$, and vanish upon integration over $\d\theta^s$. Such terms
are annihilated by $\pi_*$.)

Now comparing the basic property (\ref{edzo}) of integration over the fibers of a fibration with the definition (\ref{odz}) of
integration of an integral form, we see that
\begin{equation}\label{odox}\int_M\sigma=\int_{{\Pi} TM}\D(x,\d x)\,\omega(x,\d x)=\int_M\omega.\end{equation}
Here $\int_M\sigma$ is a Berezin integral and $\int_M\omega$ is the integral of an integral form.
This is the equivalence between the two notions of integration on a supermanifold.

The attentive reader may notice a sleight of hand in this explanation.  In discussing the integral over the fibers
of a fibration in section \ref{fibration}, we assumed that we were dealing with an ordinary integral.  But for cs manifolds, the integral
over the fibers of ${\Pi} TM\to M$ is a formal algebraic operation, as explained at the end of section \ref{pint}.  However, this algebraic
operation does have the necessary properties -- notably it transforms like an ordinary integral under a change of variables -- for the above derivation.
(In fact, locally, the $\theta$'s can be given a real structure, and the algebraic operation is equivalent to an ordinary integral.  See footnote \ref{humbug} in section
\ref{submanifolds}.
As usual, the discussion
of the equivalence between the two types of integral can be reduced to the local case by taking $\omega$ to be a sum of forms each of which is supported
in a small open set.)

\subsubsection{Integration Over Submanifolds Of Codimension $r|0$}\label{intsub}

So far we have understood that a compactly supported integral form on $M$ of codimension zero can be naturally integrated over $M$.
What can we do with an integral form of codimension $r$?

The answer to this question is that if $N\subset M$ is a supermanifold of codimension $r|0$ whose normal bundle is oriented, then an integral
form of codimension $r$ can be naturally integrated over $N$.  All that one really needs to know here is that associated to such an $N$ there is
a Poincar\'e dual $r$-form $\delta_N$, a sort of delta function $r$-form supported on $N$.  If $N$ is locally defined by vanishing of even functions $f_1\dots f_r$,
which are real-valued when the odd variables $\theta^s$ all vanish, 
 then one can define
\begin{equation}\label{oddo}\delta_N=\delta(f_1)\dots \delta(f_r)\d f_1\wedge\dots \wedge \d f_r,\end{equation}
where one orders the factors so as to agree with the orientation of the normal bundle to $N$.   Just as on a bosonic manifold, 
this formula for $\delta_N$ does not depend on the
choice of the functions $f_i$, so it makes sense globally. 

Now recall that there is a naturally defined wedge product of a differential form with an integral form.  If $\mu$ is an integral form of codimension $r$,
then $\delta_N \wedge\mu$ is an integral form of top dimension.  So we can define
\begin{equation}\label{subint}\int_N\mu=\int_M \delta_N\wedge \mu.\end{equation}

Suppose that we displace $N$ slightly to a nearby submanifold $N'$ (or in general, suppose we replace $N$ by a homologous submanifold $N'$).
In such a situation, just as for bosonic manifolds, one has
\begin{equation}\label{xon}\delta_N-\delta_{N'}=\d\tau\end{equation}
where $\tau$ is a compactly supported $r-1$-form.  So
\begin{equation}\label{subinx}\int_N\mu-\int_{N'}\mu=\int_M\d\tau\wedge \mu.\end{equation}
Now suppose that $\d\mu=0$.  Then $\d\tau\wedge\mu=\d(\tau\wedge\mu)$ so
\begin{equation}\label{binx}\int_N\mu-\int_{N'}\mu=\int_M\d(\tau\wedge\mu). \end{equation}
The supermanifold version of Stokes's theorem, to which we turn presently (eqn. (\ref{zeof})), ensures that the right hand side of (\ref{binx}) vanishes.
So if $\d\mu=0$ and $N$ is homologous to $N'$, we have
\begin{equation}\label{inx}\int_N\mu=\int_{N'}\mu,\end{equation}
just as for differential forms on an ordinary manifold.

\subsection{The Supermanifold Version Of Stokes's Theorem}\label{stokes}

If $\nu$ is an integral form of codimension 1 on a supermanifold, then $\d\nu$ is an integral form of codimension 0 and one can try to integrate it.
The most basic statement of the supermanifold version of Stokes's theorem is simply that if $\nu$ is a compactly supported integral form on $\R^{p|*q}$ of codimension 1,
then
\begin{equation}\label{ubint}\int_{\R^{p|*q}} \,\,\d \nu=0.  \end{equation}
If fact, if we write $\d=\d_0+\d_1$ with
\begin{equation}\label{bint}\d_0=\sum_{i=1}^p\d t^i\frac{\partial}{\partial t^i},~~  \d_1=\sum_{s=1}^q\d\theta^s\frac{\partial}{\partial\theta^s},\end{equation}
then 
\begin{equation}\label{nuint}\int_{\R^{p|*q}}\d_0\nu= \int_{\R^{p|*q}}\d_1\nu=0. \end{equation}
The integral of  $\d_1\nu$ over the odd variables 
vanishes because  $\d_1\nu$ is a sum of terms none of which are proportional to the product $\theta^1\dots\theta^q$ of all odd variables.    And the integral of $\d_0\nu$ over the even
variables vanishes by the ordinary bosonic version of Stokes's formula.  

We can immediately extend this to a general supermanifold $M$.  If $\nu$ is a compactly supported integral form on $M$ of codimension 1, then
\begin{equation}\label{zeof}\int_M\d\nu=0. \end{equation}
To show this, we proceed just as in the definition of the Berezin integral.  We write $\nu$ as the sum of codimension 1 integral forms $\nu_\alpha$,
each of which is supported in an open set $U_\alpha\subset M$ that is isomorphic to an open set in $\R^{p|*q}$.  So  Stokes's formula for an arbitrary $M$
follows from the special case $M=\R^{p|*q}$.

Just as in the bosonic case, a more general version of Stokes's theorem applies to a supermanifold with boundary.  First we have to define a supermanifold with boundary.
This is a little tricky and there are several ways to proceed. The simplest way to find the right definition is to first consider the
case  that everything is happening inside a supermanifold $Y$ without boundary.
  In $Y$, one considers a submanifold $N$ of codimension 1 defined by an equation $f=0$.  As usual, $f$ is required to be real when (but in general only when) the odd
  variables vanish.
For the moment, we assume that the function
$f$ is globally-defined, in which case,\footnote{The topological fact that we are using is that a codimension 1 submanifold $N_\red\subset Y_\red$ can be defined
by a globally-defined real-valued function $f_\red$ if and only if $N_\red$ is the boundary of some $M_\red\subset Y_\red$.  In one direction,  if $f_\red$ exists,
we define $M_\red$ by the condition $f_\red<0$.  In the other direction, one uses the fact that the cohomology class Poincar\'e dual to $N_\red$ vanishes if $M_\red$ exists. This
implies that the object that can always be written locally as $\delta_{N_\red}=\delta(f)\d f$ is in fact $\d\Theta(-f)$ for a globally-defined $f$.}  
roughly speaking,
we can define a compact supermanifold $M$, with boundary $N$, by the condition $f\leq 0$.  Let $\Theta(x)$ be the function of a real
variable that is 1 for $x\geq 0$ and 0 for $x<0$.  Then for any  integral form $\sigma$ on $Y$ of codimension 0 such that $\Theta(-f)\sigma$ has compact support
(we do not assume that the support
of $\sigma$ is contained in $M$), we define
\begin{equation}\label{yrt}\int_M\sigma =\int_Y\Theta(-f)\sigma.\end{equation}
Since $f$ is only real modulo nilpotents, the interpretation of this formula involves considerations such as 
$\Theta(a+b\theta^1\theta^2)=\Theta(a)+b\theta^1\theta^2\delta(a)$, where $a$ is real but $b$ need not be.  Because even nilpotent expressions like $b\theta^1\theta^2$
are neither positive nor negative (or even real, in general), it is actually the integration
formula that gives a precise meaning to the statement that $M$ is defined by the condition 
$f\leq 0$.

Why is this an interesting situation to consider?
Superstring perturbation theory provides a good example.  In that context, let $Y$
be the moduli space of super Riemann surfaces.  $Y$ is not compact -- it has noncompact ends corresponding to the infrared region --  
and one often needs to integrate over $Y$ an integral form whose behavior at infinity is delicate.
One may want to introduce an infrared regulator by restricting the integral over $Y$ to an integral over a large compact subset $M\subset Y$.
The version of Stokes's theorem that we are in the process of describing governs the boundary terms that will arise in integration by parts in this sort of situation.

With this understood, let us take the form $\sigma$ in (\ref{yrt}) to be exact, say $\sigma=\d\nu$.  Then
\begin{equation}\label{zyrt}\int_M\d\nu=\int_Y\Theta(-f)\d\nu=\int_Y\d(\Theta(-f)\,\nu)-\int_Y(\d(\Theta(-f)))\,\nu.\end{equation}
On the right hand side of (\ref{zyrt}), we can drop the exact term $\d(\Theta(-f)\nu)$, by using Stokes's theorem (\ref{zeof}) for a supermanifold without boundary.
On the other hand, $\d(\Theta(-f))=-\delta(f)\d f=-\delta_N$, where $\delta_N$ was defined in eqn. (\ref{oddo}).    So, from (\ref{subint}),
we have
\begin{equation}\label{myrt}\int_M\d\nu=\int_N \nu,\end{equation}
where $N=\partial M$ is defined by $f=0$.    This is Stokes's theorem in this situation.

This construction did not really depend on the specific choice of the function $f$.  That is because the function $\Theta(-f)$, which is all we really used, is invariant under
\begin{equation}\label{udz} f\to e^\phi f,\end{equation}
where $\phi$ is real when the odd variables vanish.  By contrast, $\Theta(-f)$ is not invariant under something like $f\to f+\alpha\beta$ where $\alpha$ and $\beta$
are odd variables (coordinates or moduli), since  $\Theta(-f-\alpha\beta)=\Theta(-f)-\alpha\beta\delta(-f)$.  So the definition of $M$ and $N$ really relied on an equivalence
class of functions $f$ modulo the relation (\ref{udz}).  We do not need $f$ to be globally-defined (though it actually is always possible to find a globally-defined $f$); 
only $\Theta(-f)$ has to be globally-defined.  It suffices to cover the
region  near the boundary of $M$ with open sets $V_a$ in each of which one is given a function $f_a$ such that 
$f_a=\exp(\phi_{ab})f_b$ in $V_a \cap V_b$, for some function $\phi_{ab}$.  Then $\Theta(-f_a)=\Theta(-f_b)$ in $V_a\cap V_b$.  This gives a function that we can call $\Theta(-f)$ that is defined
near the boundary of $M$ (that is, near the vanishing locus of any of the $f_a$) and since it is 1 in the interior of $M$ wherever it is defined (that is in $\cup_aV_a$), we can extend its definition so that it
equals 1 throughout the interior of $M$.

\subsection{Supermanifold With Boundary}\label{urtz}

Having come this far, we can give the appropriate definition of a supermanifold $M$ with boundary  without requiring $M$ to be embedded in a supermanifold $Y$ without boundary.
The only subtle point is that part of the definition of $M$ involves a function $f$, or more precisely an equivalence class (\ref{udz}) of such functions, whose vanishing defines the boundary of $M$.

We suppose that the reduced space $M_\red$ of $M$ is an ordinary manifold with boundary of dimension $p$.  
We cover $M$  with small open sets $U_\alpha$ that do not intersect its boundary, and small open sets $V_a$
that do intersect its boundary.  Each $U_\alpha$ is endowed as usual with 
local coordinates $t^1_\alpha\dots t^p_\alpha|\theta^1_\alpha\dots \theta_\alpha^q$ providing an isomorphism with an open
subset in $\R^{p|*q}$.    Each $V_a$ is likewise endowed with local coordinates $t^1_a\dots t^p_a|\theta^1_a\dots \theta^q_a$, now providing an isomorphism with an open subset of
the half-space $t^p\leq 0$ in $\R^{p|*q}$. Thus, the boundary of $M$ is always at $t^p_a=0$.  As usual, we have gluing functions in intersections $U_\alpha\cap U_\beta$, $U_\alpha\cap V_b$, and $V_a\cap V_b$.  The only
detail that is in any way special to the case of a supermanifold with boundary is that in $V_a\cap V_b$, we put
a special condition on one of the gluing functions:
we require that the gluing function defining $t^p_a$ when expressed in terms of $t^1_b\dots|\dots \theta^q_b$ takes
the form 
\begin{equation}\label{ygy}t^p_a=\exp(\phi_{ab}(t^1_b\dots|\dots\theta^q_b))t^p_b.\end{equation}
In other words, $t^p_a$ and $t^p_b$ are equivalent in the sense of  (\ref{udz}).  (In fact,  it is always possible to pick coordinates such that $\phi_{ab}=0$ and $t^p_a=t^p_b$,
though there is no particularly natural way to do this.)  

The above construction defines what we mean by a $p|q$-dimensional cs supermanifold  with boundary.  The relation (\ref{ygy}) ensures that we can consistently set $t^p_a=t^p_b=0$ and
these conditions define an ordinary  cs  supermanifold $N$ without boundary of dimension $p-1|q$.   

Integration on a supermanifold with boundary is defined in a way that should be almost obvious.  If $\sigma$ is a  compactly supported section of $\Ber(M)$ whose support is in just one of the
$U_\alpha$ or $V_a$, its integral is defined by a naive Berezin integral  (\ref{helmbo}); the integral of a general section of $\Ber(M)$ is defined with the help of a partition of unity.
The only subtlety is that if $\sigma$ has compact support in one of the boundary open sets $V_a$, then $\int_{V_a}\sigma$ is invariant under those coordinate transformations
that act on $t^p_a$ by $t^p_a\to e^\phi t^p_a$, but not under something like $t^p_a\to t^p_a+\theta_a^1\theta^2_a$.  
  This is the reason that the condition (\ref{ygy}) is part of the definition of a supermanifold with boundary.\footnote{Here is  a simple example on a half-space $t\leq 0$ in
   $\R^{1|*2}$.
  The Berezin integral $I=\int [\d t|\d\theta^1\d\theta^2]w(t)$  vanishes if $w$ is a function of $t$ only.  But if we are permitted to transform from $t$ to $t^*=t+\theta^1\theta^2$, we get
  $I^*=\int[\d t^*|\d\theta^1\d\theta^2]  w(t^*-\theta^1\theta^2)=\int [\d t^*|\d\theta^1\d\theta^2]\left(w(t^*)-\theta^1\theta^2 w'(t^*)\right)$, and now there is a boundary term at $t^*=0$. If
  $w$ is compactly supported, there is no boundary term at $t^*=-\infty$.}

Stokes's theorem for a supermanifold with boundary says that if $\nu$ is an integral form on $M$ of codimension 1 than
\begin{equation}\label{melbo}\int_M\d\nu=\int_N\nu.\end{equation}  If the gluing functions of $M$ are real analytic, so that they can continued to positive (but perhaps small) values
of $t^p_a$, then this theorem is not more general than (\ref{myrt}).  However, it holds whether or not there is such a continuation.  As usual, one first proves it by reduction to the ordinary
form of Stokes's theorem for the case that $\nu$ is compactly supported in just one of the $U_\alpha$ or $V_a$; the general case follows by using a partition of unity.

This construction is useful in superstring perturbation theory in the presence of D-branes and/or orientifold planes, since the moduli space of open and/or unoriented super Riemann
surfaces is a supermanifold with boundary in the sense just described.

\subsection{Integration On More General Submanifolds}\label{irred}

We have so far considered two types of representation of the Weyl algebra that was introduced in section \ref{informs}.
Differential forms correspond to functions on ${\Pi} TM$ with polynomial dependence on the even variables $\d\theta^i$.
They can be integrated over submanifolds of $M$ of dimension $p|0$, that is submanifolds with zero fermionic
dimension. Integral forms correspond to distributional functions on ${\Pi} TM$ with support at $\d\theta^i=0$.
They can be integrated over submanifolds with maximal fermionic dimension. 

If we want to be able to integrate over submanifolds of $M$ whose odd dimension is positive, but not maximal, we must introduce
more general representations of the Weyl algebra.  We need functions on ${\Pi} TM$ that have polynomial behavior with respect to some
of the $\d\theta^i$ and compact support with respect to others.  Let us practice with a toy example of $\R^{0|*2}$, with odd coordinates
$\theta^1$, $\theta^2$.  We can represent the Weyl algebra starting with a state $|\negthinspace\negthinspace\uparrow\negthinspace\downarrow\rangle$
that is annihilated by $\d\theta^1$ and by $\partial/\partial\d \theta^2$.  In the language of superstring perturbation theory, this choice of
``picture'' is midway between differential forms and integral forms.  

For an example of integrating a form of this type, let us take
\begin{equation}\label{guf}\omega=\theta^2\delta(\d\theta^1) \end{equation}
and try to integrate over the $0|1$-dimensional subspace $N\subset\R^{0|*2}$ defined by the equation
\begin{equation}\label{enofo} a\theta^1+b\theta^2=0,~~a,b\in\C.\end{equation}
Along $N$, we can eliminate  $\theta^2$ by $\theta^2=-(a/b)\theta^1$, $\d\theta^2=-(a/b)\d\theta^1$.  So we can parametrize
$N$ by $\theta^1$, and the integral we have to do is
\begin{equation}\label{zub}  \int_{\R^{0|*1}}\D(\theta^1,\d\theta^1)\,(-a/b)\theta^1\delta(\d\theta^1)=-\frac{a}{b}.\end{equation}
This is homogeneous in the parameters $a,b$, as it must be, since $N$ is unchanged in scaling those parameters.  
A noteworthy fact, however, is that the integral has a pole at $b=0$.  It arises because if $b=0$, the definition
of $N$ gives $\theta^1=0$, so $\d\theta^1$ is zero when restricted to $N$.  But we cannot restrict $\delta(\d\theta^1)$ to $\d\theta^1=0$.
So $\sigma$ can be integrated over a generic codimension $0|1$ submanifold of $\R^{0|*2}$, but not over every one.

More generally, given any supermanifold $M$, we consider any class of (distributional) functions on ${\Pi} TM$ with three basic properties:

(1)  For any local coordinate system $t^1\dots|\dots\theta^q$ on $M$, the class of functions is closed under the Clifford-Weyl algebra,
that is under the action of $\d x$ and $\partial/\partial \d x$, where $x$ is any of $t^1\dots|\dots\theta^q$.

(2)  The class of functions on ${\Pi} TM$ is also closed under multiplication by any function $f(t^1\dots|\dots\theta^p)$ on $M$, and under addition.

(3) Finally, we require that the given class of forms is invariant under the scaling symmetry (\ref{dromo}) of ${\Pi} TM$ that scales all fiber coordinates
by a common factor $\lambda$, and we say that a form that transforms as $\lambda^r$
under this operation  has scaling weight $r$.  

We call such objects pseudoforms  (this terminology differs slightly from the literature) or sometimes simply forms.
The conditions imply among other things that the exterior derivative \begin{equation}\label{gelb}
\d=\sum_{I=1\dots | \dots q} \d x^I\frac{\partial}{\partial x^I}  \end{equation}
 acts on functions of the given class, increasing the scaling weight by 1, and that it is possible
to multiply a function of the given class by an ordinary differential form (understood as a function on ${\Pi} TM$  with polynomial dependence on the $\d\theta$'s), increasing
its scaling weight by the degree of the differential form.

The requirement of scale-invariance of the given class of functions
implies that the support of the functions in
the space parametrized by $\d\theta^1\dots\d\theta^q$ is a conical submanifold.   For superstring perturbation theory, it seems sufficient to consider
the case that the cone is just a linear subspace -- so that we consider wavefunctions with polynomial dependence on some of the $\d\theta$'s and delta
function dependence on the others.  
If the wavefunctions are localized in $s$ variables, 
we say that 
the class of functions in question correspond to pseudoforms of  picture number $-s$.  The terminology is suggested by the usual terminology in superstring perturbation
theory.  The picture number is constant for a whole class of pseudoforms corresponding to a representation of the Clifford-Weyl algebra. 
Clearly, there are many classes of pseudoforms with the same picture number, since there are
many linear subspaces (or nonlinear cones) with the same dimension.  Some operations that change the picture number will be described in section \ref{operations}.

If a form has scaling weight $r$ and is localized with respect to $n$ $\d\theta$'s (so its picture number is $-n$), we call it a form of superdegree $m|n$, with $m=r+n$.  A simple example of a form of
superdegree $m|n$ is 
\begin{equation}\label{formtype} \omega= f(t^1\dots|\dots\theta^q) \d t^1\dots\d t^m\,\delta(\d\theta^1)\dots\delta(\d\theta^n).\end{equation}
We have used the first $m$ $\d t$'s and the first $n$ $\d\theta$'s in writing this formula.  The form $\omega$ contains $n$ delta functions of $\d\theta$'s,
and has scaling weight $m-n$, so it is indeed of superdegree $m|n$.  The exterior derivative increases the scaling weight by 1 without changing the picture number,
so it  maps forms of superdegree $m|n$ to forms of superdegree $m+1|n$.  

Given a form of superdegree $m|n$, we can try to integrate it on a submanifold $N\subset M$ of dimension $m|n$.  We say ``try'' because though there is a rather natural
operation, it is not defined for all $N$; for some choices of $N$, one will run into problems, as in the simple example that we gave above with $\R^{0|*2}$.  This should
not be too discouraging, since something similar happens
 in ordinary calculus: if one is given a $k$-form on an ordinary noncompact manifold, one can try to integrate it over $k$-dimensional submanifolds, but sometimes
 the integral will turn out to diverge.

The procedure for integration uses the fact that if $N$ is embedded in $M$, then ${\Pi} TN$ is embedded in ${\Pi} TM$.  So we can try to restrict a function $\omega(x,\d x)$
on ${\Pi} TM$ to a function on ${\Pi} TN$, which we will call by the same name. The only thing that may go wrong is that to make the restriction, we may need
to evaluate a delta function $\delta(\d\theta)$ (where $\theta$ is some linear combination of the odd coordinates) at $\d\theta=0$; this is what happened at $b=0$ in
the practice example.  
 If $\omega(x,\d x)$ is localized with respect to $n$ $\d\theta$'s, and $N$ has fermionic  dimension $n$,  then this  will not occur at a generic point on a generic $N$.
 If $N$ is such that $\omega(x,\d x)$ can be restricted to ${\Pi} TN$ without running into trouble anywhere, then the restriction is everywhere localized with respect to all
 $\d \theta$'s, and is an integral form on $N$.
 If $\omega(x,\d x)$ is $m|n$-form, then its scaling dimension is $m-n$, which ensures  that if $\omega(x,\d x)$ can indeed be restricted to give
an integral form on $N$, then the resulting integral form is a top form.   Given all this, the form $\omega(x,\d x)$ can be integrated over $N$
in the usual way.  

Suppose instead that $U\subset M$ has dimension $m+1|n$, with $\omega$ still an $m|n$-form,
and suppose that $\omega(x,\d x)$ can be restricted to $U$.  Then its support as a function of $\d\theta$
is entirely at the origin, so the restriction of $\omega(x,\d x)$ to $U$ is again 
an integral form.  But the scaling dimension of $\omega(x,\d x)$ is too small by 1 to make a top form on $U$;
rather, $\omega(x,\d x)$  is an integral form of codimension 1.      
So we cannot integrate $\omega$ over $U$, but we can integrate $\d\omega$.

Now let $U$ have boundary\footnote{Strictly speaking, to match the definition we gave in section \ref{stokes} of integration on a 
supermanifold with boundary, we should proceed
here in more steps.  We should first introduce a dimension $m+1|n$ submanifold $Y\subset M$ without boundary.  Then we let $N$ be a 
codimension one submanifold of $Y$
defined by an equation $f=0$, and we define $U$ by $f<0$.}  $N$.  Then, applying the supermanifold version of Stokes's theorem to 
the integral form $\omega$ on $U$, we have
\begin{equation}\label{zoboxx}\int_U\d\omega =\int_N\omega. \end{equation}
It follows from this that if $\d\omega=0$, then $\int_N\omega$ is invariant under small displacements of $N\subset M$ and 
more generally under a certain class of allowed homologies.    

\subsubsection{Another Example}\label{dolfott}

We conclude by describing another entertaining example.  In $\R^{1|1}$, with coordinates $t|\theta$, we consider the form
\begin{equation}\label{ifo} \omega = \theta\, \d t\, \delta'(\d\theta),  \end{equation}
This form has scaling degree $-1$ and picture number $-1$.  So it is a $0|1$-form, and we can hope to integrate it over a submanifold $N$
of $\R^{1|1}$ of dimension $0|1$.  We define a suitable $N$ by the equation $t=\alpha\theta$, with $\alpha$ an odd parameter.  Upon restriction to
$N$, we have $\d t =-\alpha \d\theta$, so $\omega=-\theta\alpha \d\theta \delta'(\d\theta)=\theta\alpha \delta(\d\theta)$.  We used the fact that for
an even variable $y$, $y\delta'(y)=-\delta(y)$.  So
\begin{equation}\int_N\omega =\int \D(\theta,\d\theta) \theta\alpha \delta(\d\theta)=\alpha.\end{equation}

\section{More Operations On Forms}\label{operations}

\subsection{Wedge Products And Contractions}\label{wedgep}

Here we will describe some interesting operations on pseudoforms, largely following \cite{Beltwo,Belthree}.     Some of this material may be
 useful background for superstring perturbation theory. (On some points whose relevance is not immediately clear, we provide only references
 to the literature.)

One basic operation is the wedge product with a one-form.  If $\alpha$ is a one-form on $M$, we define an operator $\e_\alpha$
that acts on pseudoforms by multiplication by the corresponding function $\alpha(x,\d x)$ on ${\Pi} TM$:
\begin{equation}\label{obz}[\e_\alpha\omega](x,\d x) =\alpha(x,\d x) \omega(x,\d x). \end{equation}
This operation increases the degree by 1 and does not change the picture number.  The statistics of $\e_\alpha$ are the same as 
those of $\alpha$.  So $\e_\alpha$ is odd if $\alpha=\d t$ with $t$ an even variable, but $\e_\alpha$ is even if $\alpha=\d\theta$ where $\theta$ is odd.

For a very simple operation that can change the picture number, we define an operator $\delta(\e_\alpha)$ as multiplication by $\delta(\alpha(x,\d x))$:
\begin{equation}\label{nobz}[\delta(\e_\alpha)\omega](x,\d x)=\delta(\alpha(x,\d x))\omega(x,\d x). \end{equation}
These operators obey
\begin{equation}\label{nobozx}\e_\alpha\delta(\e_\alpha)=\delta(\e_\alpha)\e_\alpha=0.\end{equation}
If $\alpha$ is an odd one-form such as $\d t$, then $\e_\alpha$ is an odd variable.  Since an odd variable is its own delta function, we have simply
\begin{equation}\label{ozo}\delta(\e_\alpha)=\e_\alpha. \end{equation}
In particular, in this case $\delta(\e_\alpha)$ is fermionic, does not change the class of a form, and maps an $m|n$-form to an $m+1|n$-form.
But if $\alpha$ is even, for instance $\alpha = \d\theta$, then multiplication by $\delta(\alpha)=\delta(\d\theta)$ is the most simple example
of an operator that changes the class of a form.  It maps an $m|n$-form to an $m|n+1$-form.  The operator $\delta(\e_\alpha)$ is odd regardless of whether
$\alpha$ is even or odd, though for even $\alpha$ this is subtle; see eqn. (\ref{obort}).   For even $\alpha$, the operator $\delta(\e_\alpha)$
is not defined on all classes of pseudoform, since an object with support at $\d\theta=0$ cannot be multiplied by $\delta(\d\theta)$.  

If $M$ is a supermanifold of dimension $p|q$, and $\gamma^1\dots\gamma^q|\zeta^1\dots\zeta^p$ is a basis of even and odd one-forms,
then the operation
\begin{equation}\label{frobz} f\to \delta(\e_{\gamma^1})\dots \delta(\e_{\gamma^q})\delta(\e_{\zeta^1})\dots \delta(\e_{\zeta^p}) f \end{equation}
supplies all the missing factors of $\d t$ and $\delta(\d\theta)$ and
maps a function $f$ to an integral form of top degree.

The dual operation to multiplication by a one-form is contraction with a vector field on $M$.  For a vector field 
\begin{equation}\label{ork}V=\sum_{I=1\dots|\dots q} V^I\frac{\partial}{\partial x^I},\end{equation}
 we define the contraction operator
\begin{equation}\label{dork}\i_V=\sum_{I=1\dots|\dots q} V^I\frac{\partial}{\partial \d x^I}\end{equation}
This operator has the opposite statistics to $V$.  It maps a form of superdegree $m|n$ to one of superdegree $m-1|n$.
It is again useful to introduce delta function operators that will obey
\begin{equation}\label{orto}\i_V\delta(\i_V)=\delta(\i_V)\i_V=0.\end{equation}
Here at first we treat separately the cases of an even vector field $v$ or an odd vector field $\nu$.
If $v$ is an even vector field, so that $\i_v$ is odd, then again the definition of $\delta(\i_v)$ is obvious:
\begin{equation}\label{zoto}\delta(\i_v)=\i_v.  \end{equation}

It is less obvious how to define $\delta(\i_\nu)$ for an odd vector field $\nu$.  However, the appropriate definition has been given in \cite{Beltwo}.
If $\nu$ is an odd vector field on $M$, it can be viewed as a section of ${\Pi} TM$.  And hence, for $u$ an even scalar, $u\nu$ is a section of ${\Pi} TM$
and it makes sense to act on ${\Pi} TM$ by shifting the fiber coordinates $\d x$ by $\d x \to \d x+ u\nu$.  This makes possible the definition
\begin{equation}\label{ohon}[\delta(\i_\nu)\omega](x,\d x)= \int[\d u]\,\, \omega(x,\d x + u\nu),\end{equation}
(for $\omega$ in a suitable class of pseudoforms)
which can be seen to satisfy (\ref{orto}).
This operation maps a form of superdegree $m|n$ to a form of superdegree $m|n-1$.  The operator $\delta(\i_\nu)$ is odd, since the integration form $[\d u]$ is odd.
Thus, for example, if $\nu$ and $\nu'$ are two odd vector fields, we have
\begin{equation}\label{bohon}[\delta(\i_\nu)\delta(\i_{\nu'})\omega](x,\d x)=\int [\d u \,\d u']\,\,\omega(x,\d x+ u\nu+u'\nu'),\end{equation}
and this is odd in $\nu$ and $\nu'$ since $[\d u \,\d u']=-[\d u'\,\d u]$.  Thus $\delta(\i_V)$ is odd regardless of whether $V$ is even or odd.

A little thought shows that actually we can define the delta function operation in the same way also for an even vector field $v$.  If $v$ is an
even vector field and $\eta$ is an odd constant, then $\eta v$ is an odd vector field and thus again a section of ${\Pi} TM$.  So we can define
\begin{equation}\label{hono}[\delta(\i_v)\omega](x,\d x)=\int [\d \eta]\,\omega(x,\d x+ \eta v), \end{equation}
and this is equivalent to the previous definition.  

Now suppose that we are given a collection of even and odd vector fields $v_1\dots v_m|\nu_1\dots \nu_n$.   Acting with the whole product of
delta functions 
\begin{equation}\label{zormo}\delta(\i_{v_1})\dots\delta(\i_{v_m})\delta(\i_{\nu_1})\dots\delta(\i_{\nu_n}) \end{equation}
we get an operator that integrates over the $m|n$-dimensional subspace of the fibers of ${\Pi} TM$ that is generated by $v_1\dots|\dots\nu_n$.
Explicitly, to act with the product of delta function operators, we  introduce $n$ even and $m$ odd
integration variables $u^1\dots u^n$ and $\eta^1\dots \eta^m$
and perform the Berezin integral
\begin{equation}\label{ormo}\int[\d u^1\dots|\dots \d\eta^m]\,\,\omega\left(x,\d x+\sum u^i\nu_i+\sum \eta^j v_j\right),\end{equation}
assuming that $\omega$ is a form of an appropriate type so that this integral makes sense.
Let  us denote the integral as $\sigma(v_1\dots|\dots \nu_n)$.  Actually, $\sigma$ is a function of $x$ and possibly the fiber variables in ${\Pi} TM$
that we have not integrated over; it depends on the values at $x$ of the vector fields $v_1\dots|\dots\nu_n$.  To simplify the notation, we denote $\sigma$
simply as $\sigma(v_1\dots|\dots\nu_n)$.

The vector fields $v_1\dots |\dots \nu_n$ span a subbundle $V_{m|n}$ of the tangent bundle $TM$.  Reversing the statistics gives
a subbundle ${\Pi} V_{m|n}$ of ${\Pi} TM$.  The integral in (\ref{ormo}) is an integral over ${\Pi} V_{m|n}$, but it is not quite true that the integral
$\sigma(v_1\dots|\dots\nu_m)$
depends on the chosen vector fields only via  the subspace they generate.  If we replace $v_1\dots|\dots \nu_n$ by another collection of vector fields
$v'_1\dots|\dots \nu'_n$ that span the same subspace,  with
\begin{equation}\label{nodexo}\begin{pmatrix}v_1'\cr v_2'\cr \vdots \cr - \cr \vdots \cr \nu'_n\end{pmatrix} =
W \begin{pmatrix}v_1\cr v_2\cr \vdots \cr - \cr \vdots \cr \nu_n\end{pmatrix},\end{equation}
for some linear automorphism $W$ of ${\Pi} V_{m|n}$, then we can compensate for this in (\ref{ormo}) by redefining the integration
variables $u^1\dots| \dots \eta^n$ by a dual linear automorphism, but this will change the integration measure by a factor $\Ber(W)$.
So we get
\begin{equation}\label{zormon} \sigma(v'_1\dots|\dots \eta'_n)=\Ber(W)\sigma(v_1\dots|\dots\eta_n). \end{equation}

A special case of this is that $m|n$ equals the dimension $p|q$ of $M$.  In this case, $v_1\dots|\dots \eta_n$ is a basis of the tangent space to $M$.
A function depending on such a basis and obeying (\ref{zormon}) is a section of $\Ber(M)$, as explained in section \ref{relevance}.  The integral 
in (\ref{nodexo}) is simply an integral over the fibers of ${\Pi} TM\to M$, and what we have arrived at is an operation already described in section \ref{dirred}:
the map from functions on ${\Pi} TM$ to sections of  $\Ber(M)$, by integrating over the fibers of ${\Pi} TM$.

If on the other hand $m|n$ does not coincide with $p|q$, then we have described something more general. For some choices of $\omega(x,\d x)$, a function
$\sigma(v_1\dots|\dots\nu_n)$ obtained by integrating over an $m|n$-dimensional subbundle of ${\Pi} TM$ may still depend on the other fiber coordinates.
However, if $\omega$ is a pseudoform of degree $m|n$ in the sense described in section \ref{irred}, then (for a given choice of vector fields $v_1\dots|\dots\eta_n$)
$\sigma$ will be a function on $M$, and not a more general function on ${\Pi} TM$.  
In this case, $\sigma(v_1\dots|\dots\nu_n)$ is a differential form on $M$ of degree $m|n$, in the language of  \cite{VZ,VZtwo}, as reviewed in \cite{Voronov}.
Such an object is by definition a function of a point $x\in M$ that  depends on  $m$ even and $n$
odd vectors in the tangent space to $x$ in $M$, with the restriction  (\ref{zormon}), and also obeys a certain fundamental relation 
that is described on p. 57 of \cite{Voronov}.  This relation
is automatically satisfied when $\sigma$ is defined by an integral (\ref{ormo}).  The fundamental relation might be important for some sort
of string or brane actions on supermanifolds.

\subsection{Picture-Changing}\label{picturechanging}

For a vector field $V$, the operator $\delta(\i_V)$ defined in section \ref{operations} is not invariant under multiplying $V$ by a constant, $V\to \lambda V$.
Rather this operation rescales $\delta(\i_V)$ by $\lambda$ or $\lambda^{-1}$, depending on the statistics of $V$.  

However, for the case of an odd vector field $\nu$, there is a natural ``picture-changing'' operation, defined in \cite{Beltwo}, which is invariant under $\nu\to\lambda\nu$
and only depends on the $0|1$-dimensional group of automorphisms of $M$ generated by $\nu$.  Let us call this group $F$; it is isomorphic to $\R^{0|1}$ (or
$\R^{0|*1}$).  
The $F$ action on $M$ corresponds to a map $m:F\times M\to M$.  Given a pseudoform $\omega(x,\d x)$ on $M$, we can pull it back to a form $m^*(\omega)$
on $F\times M$.  Then we have a projection $\pi:F\times M\to M$ that forgets the first factor.  Integrating over the fibers of $\pi$, we get again
a pseudoform on $M$.  So this gives an operation $\Gamma_\nu=\pi_*m^*$ on pseudoforms that (because we carry out one odd integration in integrating over the fibers of $\pi$)
maps an $m|n$ form to an $m|n-1$ form.    

To make this explicit, 
we parametrize
$\R^{0|1}$ by an odd variable $\tau$, and write the action of the group on $M$ as 
\begin{equation}\label{zolk}x^I\to \exp(\tau\nu^J\partial/\partial x^J) x^I= x^I+\tau\nu^I.\end{equation}
As usual, this expansion stops quickly since $\tau^2=0$.  
The pseudoform $m^*(\omega)$ on $F\times M$ is simply \begin{equation}\label{peneo}\omega(x+\tau\nu,\d(x+\tau\nu))=\omega\left(x+\tau\nu,
\d x+\d\tau\, \nu - \tau \frac{\partial\nu}{\partial x^A}\d x^A\right).\end{equation}
And integration over the fibers means integrating over $\tau$ and $\d\tau$ with the natural measure $\D(\tau,\d\tau)$.  So the picture-changing operator
$\Gamma_\nu$ is defined by
\begin{equation}\label{zobz}[\Gamma_\nu\omega](x,\d x)=\int \D(\tau,\d\tau) \,\omega(x+\tau \nu, \d(x+\tau\nu)).\end{equation}
As explained in \cite{Beltwo,Belthree}, this is an abstract version of the picture-changing operator of superstring perturbation theory.
It is shown there that
\begin{equation}\label{oboz}\Gamma_\nu=\frac{1}{2}\left(\delta(\i_\nu)\LL_\nu+\LL_\nu\delta(\i_\nu)\right), \end{equation}
where $\LL_\nu$ is the Lie derivative,
\begin{equation}\label{boz}\LL_\nu=\d\i_\nu+\i_\nu\d.\end{equation}
The exterior derivative $\d$ commutes with $m^*$ and $\pi_*$, so it commutes with $\Gamma_\nu$.  Additional interesting facts can be found
 in \cite{Beltwo}.     
                
\section{Complex and Smooth Supermanifolds}\label{bicomplex}

\subsection{First Orientation}\label{orient}

In this concluding section, we will compare complex supermanifolds to smooth ones.  The relevance for superstring perturbation theory will become clear.

For a first orientation to the problem, suppose that $X$ is an ordinary 
complex manifold, of dimension $a$, with local holomorphic coordinates
$z^1\dots z^a$.   $X$ can always be viewed
as a smooth manifold of dimension $2a$, local real coordinates being 
$t^i=\mathrm{Re}\,z^i=(z^i+\bar z^i)/2$ and $t^{a+i}=\mathrm{Im}\,z^i=(z^i-\bar z^i)/2\sqrt{-1}$, with $i=1\dots a$.

We can do the same thing on a supermanifold $X$ -- and this is often done in the literature --
 if we are allowed to take the complex conjugates (and therefore the real and imaginary
 parts) of both even and odd variables.  This would
turn a complex manifold of dimension $a|b$ into a real supermanifold of dimension $2a|2b$.
But for superstring perturbation theory,
that is not what one wants to do, since we are never allowed to take the complex 
conjugate of an odd variable.\footnote{And consequently, unless we are presented with a splitting, 
so that we know which even variables are ``purely bosonic,'' 
we cannot take
the complex conjugate of an even variable either.}  Instead a typical
thing that we want to do is to relate $X$ to a cs supermanifold of dimension $2a|b$, 
with no doubling of the fermionic dimension.  

For  example, let $X$ be the worldsheet of a heterotic string.  From a holomorphic point of view,  
$X$ is a complex supermanifold of dimension $1|1$ with
local coordinates $z|\theta$.  (Holomorphically, a heterotic string worldsheet is
not a generic complex supermanifold of dimension $1|1$; it has the additional
structure of a super Riemann surface. The details,
which are reviewed in  \cite{supersurf},  need not concern us here.)  
But in heterotic string theory, there are antiholomorphic as well as holomorphic
worldsheet fields, and the antiholomorphic dimension is $1|0$.  
So we want to be able to view $X$ as a smooth supermanifold of dimension $2|1$.  
In what sense can we do this?

The question is trickier than one might at first think.  We will consider primarily two points of view, which generalize the following
considerations in conformal field theory on an ordinary Riemann surface.  One typically considers correlation functions which are neither holomorphic nor
antiholomorphic.  The expectation value of a product of operators $\Phi_1\dots\Phi_s$ is often written
\begin{equation}\label{mendgo} \langle\Phi_1(\bar z_1;z_1)\dots \Phi_s(\bar z_s;z_s)\rangle.\end{equation}
There are two contrasting points of view about this formula: 

(1) $\bar z$ is really the complex conjugate of $z$.  Denoting an operator as $\Phi(\bar z;z)$ is merely a way of saying that $\Phi$ is an (operator-valued)
function on $\Sigma$ that is neither holomorphic nor antiholomorphic.  

(2) A Riemann surface $\Sigma$ is a real-analytic two-manifold and as such it can be analytically continued and viewed as a real
slice in a two-dimensional complex manifold. (How to do this concretely is explained in section \ref{altpoint}.)   When this is done, $\bar z$ and $z$ become independent complex variables; to emphasize
this, we write $\t z$ instead of $\bar z$.  The correlation functions are likewise real analytic (away from singularities when distinct points collide)
so they can be analytically continued to holomorphic functions 
\begin{equation}\label{mendgox} \langle\Phi_1(\t z_1;z_1)\dots \Phi_s(\t z_s;z_s)\rangle\end{equation}
 of independent complex variables $z_i$ and $\t z_i$. To be more precise, these functions are holomorphic when $\t z_i$ is sufficiently close to $\bar z_i$.
 Setting $\t z_i=\bar z_i$, we get the usual correlation functions on $\Sigma$.  The notation (\ref{mendgo}) is a shorthand for all this.
 
In this section, we will attempt to generalize both points of view to supermanifolds.  The first point of view is perhaps more obvious, but the second
point of view seems to be more robust.  

\subsection{Complex Supermanifold As A Smooth Supermanifold}\label{bsc}

Let $X$ be a complex manifold of dimension $a|b$.  We would like to view $X$ as a smooth cs supermanifold of dimension $2a|b$.
We will first explain what appears to be the best that one can do, and then explain why the construction is not completely natural.

First of all, if $Y$ is an ordinary complex manifold of complex dimension $a$, then the complex conjugate of $Y$, which we denote $\bar Y$, is defined to be the same manifold with opposite
complex structure.  So holomorphic functions on $Y$ are antiholomorphic functions on $\bar Y$, and vice-versa.  Concretely, if one covers $Y$ by open
sets $U_\alpha$ with local holomorphic coordinates $z^i_\alpha$, $i=1\dots a$  and gluing maps
\begin{equation}\label{torkety} z^i_\alpha=f_{\alpha\beta}^i(z^1_\beta\dots z^a_\beta),\end{equation}
then $\bar Y$ is covered by the same open sets $U_\alpha$ with local holomorphic coordinates $\tilde z^i_\alpha=\bar{z^i_\alpha}$ and gluing maps
\begin{equation}\label{norkety} \t z^i_\alpha=\bar f^i_{\alpha\beta}(\t  z^1_\beta\dots\t z^a_\beta).\end{equation}
We recall that the definition of the function $\bar f$ is such that $\bar f(\bar w)=\bar{f(w)}$, so that the above relations are consistent with 
$\t z^i_\alpha=\bar{z^i_\alpha}$.

Now let $X$ be a complex supermanifold of dimension $a|b$.  We cover $X$ by open sets $U_\alpha$ and as usual describe $X$ by local
holomorphic coordinates $z^1_\alpha\dots|\dots\theta^b_\alpha$ 
with holomorphic gluing laws:
\begin{align}\label{borrox} z^i_\alpha & = f_{\alpha\beta}^i(z^1_\beta\dots|\dots \theta^b_\beta) \cr
                                                      \theta^s_\alpha& = \psi_{\alpha\beta}^s(z^1_\beta\dots|\dots\theta^b_\beta).\end{align}
$X$ also has a reduced space $X_\red$, which is an ordinary complex manifold of dimension $a$.  Its gluing relations are obtained
from those in (\ref{borrox}) by setting all odd variables -- both the $\theta$'s and the possible odd moduli of $X$ -- to zero.      (Why the
odd variables must be set to zero to define the reduced space was explained in section \ref{families}.)     
Gluing laws for $\bar X_\red$, the complex conjugate of $X_\red$, are obtained by setting the odd variables to zero and complex-conjugating:
 \begin{equation}\label{borroxx} \t z^i_\alpha  =\bar f_{\alpha\beta}^i(\t z^1_\beta\dots|0\dots 0).\end{equation}  
This ensures that 
\begin{equation}\label{onrel}\t f_{\alpha\beta}^i(\bar z^1_\beta\dots\bar z^a_\beta)=\bar{f_{\alpha\beta}^i(z^1_\beta\dots z^a_\beta|0\dots0)}.\end{equation} 

Now we introduce $2a$ real coordinates $t^1\dots t^{2a}$:
\begin{align}\label{dorsko} z^i_\alpha& = t^i_\alpha+\sqrt{-1}\, t^{i+a}_\alpha \cr
                                         \t z^i_\alpha& = t^i_\alpha -\sqrt{-1}\,t^{i+a}_\alpha. \end{align}
By virtue of (\ref{onrel}), the gluing relations (\ref{borrox}) and (\ref{borroxx}) are compatible with reality of $t^1\dots t^{2a}$ when all odd variables (including odd moduli) vanish.
So when regarded as gluing relations for the whole collection of variables $t^1\dots t^{2a}|\theta^1\dots\theta^b$, these formulas
define a smooth manifold $X_{\cs}$ of dimension $2a|b$.  In particular, the odd coordinates of $X_\cs$ are the same as those of $X$.

Starting with a complex supermanifold $X$, we have defined a smooth supermanifold $X_\cs$ on which it makes sense to discuss both holomorphic functions (functions of $z^1_\alpha\dots|\dots
\theta^b_\alpha$) and antiholomorphic functions (functions of $\t z_\alpha^1\dots \t z^a_\alpha$).  This corresponds to point of view (1) of section \ref{orient}.

\subsubsection{Critique}\label{critique}

Though this construction is valid as far as it goes, there is a flaw: the passage from $X$ to $X_\cs$ is not as natural as one would like.    It depends in a subtle way on the specific
gluing construction.

In fact, if we had a completely natural way to transform a complex supermanifold $X$ to a smooth supermanifold $X_\cs$, it would follow that for any isomorphism $\varphi:X\cong Y$
between complex supermanifolds, we would get a corresponding isomorphism $\varphi_\cs:X_\cs\to Y_\cs$.  The isomorphisms $\varphi_\cs$ would obey the same algebraic
relations as the $\varphi$.  So in particular, setting $Y=X$, the supergroup $\G$ of automorphisms of $X$ would act on $X_\cs$.

To see that this is a problem, suppose that $X=\CP^{1|1}$, with homogeneous coordinates $u,v|\theta$.  This example simply corresponds to a genus 0 worldsheet
of the heterotic string.  The automorphism supergroup is\footnote{If $\CP^{1|1}$ is viewed as a super Riemann surface, its automorphism group is reduced from $\mathrm{PGL}(2|1)$ to 
$\mathrm{OSp}(1|2)$.
This does not really affect the discussion in the text.} $\G=\mathrm{PGL}(2|1)$, acting by linear transformations
of the homogeneous coordinates.  To promote $X$ to a smooth supermanifold, we would want to introduce antiholomorphic homogeneous coordinates $\t u, \t v$, which roughly
speaking are complex conjugates of $u,v$, but of course we introduce no corresponding odd variable $\t\theta$.  There is no way for $\mathrm{PGL}(2|1)$ to act on the pair $\t u,\t v$,
so the passage from $X$ to $X_\cs$ cannot be completely natural.  

Let us see what happens if we study this example with the gluing construction.  We can cover $X$ with an open set $U_1$ in which $u\not=0$ and a second set $U_2$ in which $v\not=0$.
In $U_1$, we take coordinates $z_1=v/u,$ $\zeta_1=\theta/u$, and in $U_2$, we set $z_1=u/v$, $\zeta_2=\theta/v$.
Following the above recipe, the gluing laws of $X_\cs$ are
\begin{align}\label{irvo} z_2&= \frac{1}{z_1}\cr \zeta_2&=\frac{\zeta_1}{z_1} \cr\t z_2& =\frac{1}{\t z_1}.\end{align}

In the starting point, without changing anything else, we could have replaced $z_1$ by $z_1'=z_1+\alpha\zeta$, where $\alpha$ is an odd parameter.
This is an equally valid starting point, and  the above recipe gives a smooth supermanifold $X'_\cs$ with
\begin{align}\label{irvox} z_2& =\frac{1}{z_1'}+\frac{\alpha\zeta_1}{(z_1')^2}\cr \zeta_2&=\frac{\zeta_1}{z_1'} \cr\t z_2& =\frac{1}{\t z_1}.\end{align}  Note that the gluing
law for $\t z_1$ and $\t z_2$ is unchanged, since to define it, we are supposed to first set the odd variables to zero.

In fact, $X'_\cs$ is isomorphic to $X_\cs$ by an isomorphism that maps  holomorphic functions to holomorphic functions and antiholomorphic functions to antiholomorphic functions.
The problem is that there are multiple equally natural isomorphisms that do this.  We could transform the gluing formulas (\ref{irvox}) back into (\ref{irvo})
by replacing $z_1'$ by $z_1''=z_1'-\alpha\zeta_1$ (which happens to be the same as $z_1$) or by replacing $z_2$ with $z_2'=z_2-\alpha\zeta_2z_2$.  

So $X_\cs$ is unique up to isomorphism, but not up to a unique isomorphism.    There is no good way to pick a particular isomorphism.

The author suspects that it is better to develop a different approach, following point of view (2) from section \ref{orient}.  In fact, in the process, we will get a new
understanding of the smooth supermanifold  $X_\cs$ that was defined above.

As a preliminary, we will describe smooth submanifolds of a complex supermanifold.  Then we return to our theme in section \ref{altpoint}.

\subsection{Submanifolds Of A Complex Supermanifold}\label{zmon}

So far we have found it difficult to give a completely natural notion of a not necessarily holomorphic function on a complex supermanifold $X$.

A function on $X$ would be a map from $X$ to $\C$, or perhaps to some ring generated over $\C$ by odd elements. 

Maps in the opposite direction behave much better.  Thus, instead of a map from $X$ to $\C$, let us consider a map from some
smooth supermanifold $M$ to $X$. 

There is no problem at all in defining what we mean by a smooth map  $\phi:M\to X$.  Locally, such a map
 expresses the local coordinates $z^1_\alpha\dots|\dots\theta^b_\alpha$ of $X$ as smooth functions  of local coordinates $t^1_\tau\dots|\dots \eta^s_\tau$ of $M$.
 (To compare the descriptions in different local coordinate systems, one just asks that the image in $X$ of a given point in $M$ should not depend
 on the coordinates used, on either $M$ or $X$.)
 Moreover, there is no problem in deciding whether such a map is an embedding.
 We require that the map of reduced spaces $\phi_\red:M_\red\to X_\red$ is
 an embedding, and that the differential of the map $\phi$ in the odd directions is sufficiently generic.\footnote{If $X_\cs$
 is a smooth supermanifold associated to $X$ as in section \ref{bsc}  (the following definition does not depend on the precise
 construction of $X_\cs$) with local coordinates $u^1_\alpha\dots|\dots
 \theta^b_\alpha$, then one requires that at least one of the maximum rank minors of 
 the matrix of derivatives $\partial(u^1_\alpha\dots|\dots\theta^b_\alpha)/\partial(t^1_\tau\dots|\dots \eta^s_\tau)$ should
 have  nonzero Berezinian. This echoes the condition for a smooth map of ordinary manifolds to be a local embedding.}  If $\phi:M\to X$
 is an embedding, we call $M$ a smooth submanifold (or subsupermanifold) of $X$.
  
 There are many natural examples of such smooth submanifolds.  In fact, we 
 can adapt something  explained in  section \ref{submanifolds}.  Let $N_\red$ be any submanifold
 of the reduced space $X_\red$ of $X$.  We can view $X$ as a smooth supermanifold  by following the construction of section \ref{bsc} 
 (for the present purpose, it does not matter
 that this construction is slightly unnatural).   Then as explained in section \ref{submanifolds}, to the submanifold $N_\red\subset X_\red$,
 we can associate a submanifold $N\subset X$ not quite uniquely, but in a way that is unique up to homology (up to infinitesimal wiggling of the fermionic
 directions).   
 
 This gives an abundant source of smooth submanifolds of a complex supermanifold,
 and the construction can be further generalized by thickening $N_\red$ in only some of the
 fermionic directions.    As an application, we will generalize to a complex supermanifold 
 the notion of the periods of a holomorphic differential form of top dimension on
an ordinary complex manifold.

\subsubsection{Periods On A Complex Supermanifold}\label{periods}

If $Y$ is an ordinary complex manifold of complex dimension $a$, then a top-dimensional holomorphic form $\sigma$ on $Y$ is in particular a closed
differential form of degree $a$.  If $N$ is an oriented middle-dimensional cycle in $Y$, and thus of real dimension $a$,
 we can define the period $\int_N\sigma$, which only depends on
the homology class of $N$.   

To generalize this to a complex supermanifold $X$, we need to define the holomorphic analog of the Berezinian line bundle, which we introduced in the smooth
case in section \ref{secber}.  
The cotangent bundle of $X$ in the holomorphic sense\footnote{It seems that, just as $X$ does not have a completely natural structure of smooth supermanifold,
it does not have a completely natural tangent or cotangent bundle except in the holomorphic sense.}    is a holomorphic bundle $T^*X$ of rank $a|b$.
We define a holomorphic version of the Berezinian of $X$, which we will call $\mathit{Ber}(X)$, by imitating the definition
used for smooth supermanifolds.    To any local  system $T=z^1\dots|\dots\theta^b$ of holomorphic coordinates on $M$, we associate
a local holomorphic section of $\mathit{Ber}(X)$ that we denote $[\d z^1\dots|\dots\d\theta^b]$.  If $\tilde T=\t z^1\dots |\dots \t\theta^b$ is a second local holomorphic coordinate
system, then we relate the two sections of $\mathit{Ber}(X)$ by the formula (\ref{hockey}).

Now we want to ask in what sense a holomorphic section $\sigma$ of $\mathit{Ber}(X)$ can be integrated.  For this, one approach is to view $X$ as a smooth supermanifold
$X_\cs$ 
of dimension $2a|b$.  From that point of view, $\sigma$ corresponds to an integral form on $X_\cs$ of codimension $a$, which moreover is closed, $\d\sigma=0$.
The map from a section of $\mathit{Ber}(X)$ to an integral form takes $[\d z^1\dots |\dots\d\theta^b]$ to the integral form $\d z^1\dots \d z^a\,\delta(\d\theta^1)\dots
\delta(\d\theta^b)$.
This is of codimension $a$ in the real sense as $\d\t{z^1}\dots\d\t{z^a}$ are missing.

If $N_\red\subset X_\red$ is of middle dimension, then the corresponding cycle $N\subset X$ is of real codimension $a|0$.  So (given an appropriate
orientation condition), there is a natural integral $\int_N\sigma$.  This only depends on the homology class of $N_\red$.

So  $\sigma\in H^0(X,\mathit{Ber}(X)) $ defines a linear function on the middle-dimensional homology of $X_\red$.  By ordinary topology, this means
that $\sigma$ determines a class $[\sigma]$ in the middle-dimensional complex-valued cohomology $H^a(X_\red,\C)$.  For example, if $X$ is a super Riemann
surface, then a holomorphic section of $\mathit{Ber}(X)$ has periods associated to ordinary $A$-cycles and $B$-cycles in the ordinary
Riemann surface  $X_\red$, and defines a class in $H^1(X_\red,\C)$.  This enables one to define the super period matrix of
a super Riemann surface, though it is tricky to show that the super period matrix is symmetric. 

\subsection{Alternative Point Of View}\label{altpoint}

Having completed the preliminaries of section \ref{zmon},  we will now generalize viewpoint (2) of section \ref{orient} to supermanifolds.  

First let us spell out more precisely what viewpoint (2) means, again taking Riemann surfaces as an example.  For $\Sigma$ a Riemann surface, let $\Sigma_R$ be
a copy of $\Sigma$ and let $\Sigma_L$ be a copy of $\Sigma$ with opposite complex structure.  Then $\Sigma_L\times \Sigma_R$ is a complex
manifold of complex dimension 2, and $\Sigma$ is naturally embedded in $\Sigma_L\times \Sigma_R$ as the diagonal.  If we view
$\Sigma$ is a real analytic manifold of dimension 2, then we can regard $\Sigma_L\times \Sigma_R$ as a complexification
of $\Sigma$.  

In the last paragraph, we do not literally need to take $\Sigma$ to be the diagonal in $\Sigma_L\times \Sigma_R$.  $\Sigma$ can be
any real-analytic submanifold of $\Sigma_L\times \Sigma_R$ that is sufficiently close to the diagonal.  $\Sigma_L\times \Sigma_R$
can be understood as a complexification of any such $\Sigma$.  One can study conformal field theory on any such $\Sigma$ and its
content is essentially independent of $\Sigma$ since the correlation functions can anyway be analytically 
continued to holomorphic functions
on an open
set in $\Sigma_L\times \Sigma_R$.    Restricted to any such $\Sigma$, the correlation functions $ \langle\Phi_1(\t z_1;z_1)\dots \Phi_s(\t z_s;z_s)\rangle$
have only the usual CFT singularities when pairs of points coincide.

By a ``holomorphic function on $\Sigma$,'' we mean the restriction
to $\Sigma$ of a holomorphic function on $\Sigma_R$, and similarly an ``antiholomorphic function on $\Sigma$'' is the restriction to $\Sigma$
of an antiholomorphic function on $\Sigma_R$.  Unless $\Sigma$ is actually the diagonal in $\Sigma_L\times \Sigma_R$, it is not true
that a antiholomorphic function on $\Sigma$ is the complex conjugate of a holomorphic function.  This is analogous to the situation encountered in section
\ref{bsc} when we associated a smooth supermanifold $X_\cs$ to a complex supermanifold $X$: the antiholomorphic coordinate $\t z$ is
not quite the complex conjugate of $z$; it has this interpretation only modulo the odd variables.

We can now reinterpret the object $X_\cs$ defined in section \ref{bsc}.  Let $X_R$ be a copy of $X$ and let $X_L$ be $\bar X_\red$, that is, $X_L$ is the reduced space $X_\red$
with complex structure reversed.  The formula (\ref{dorsko}) enabled us to define a smooth supermanifold $X_\cs$ with local coordinates
$t^1_\alpha\dots t^{2a}_\alpha|\theta^1_\alpha\dots \theta^b_\alpha$.  However, bearing in mind that the odd coordinates
$\theta^i_\alpha$ of $X_R$ are the same as the odd coordinates of $X_\cs$, we can read the first line of eqn. (\ref{dorsko}) as defining
a continuous map from $X_\cs$ to $X_R$.  And the second line of eqn. (\ref{dorsko}) similarly defines a continuous map from $X_\cs$ to $X_L$.
Altogether what we have is a smooth supermanifold $X_\cs$ with an embedding $X_\cs\hookrightarrow X_L\times X_R$.

The reduced space of $X_\cs$, moreover, is the diagonal in the reduced space of $X_L\times X_R$.  And we can interpret $X_L\times X_R$
as the complexification of $X_\cs$.  (This complexification is defined by interpreting all local coordinates $t^1_\alpha \dots t^{2a}_\alpha|\theta^1_\alpha
\dots \theta^b_\alpha$ of $X_\cs$ as independent complex variables.  From eqn. (\ref{dorsko}), this just means that 
$z^i_\alpha$ and $\t z^i_\alpha$ are independent complex variables, as is appropriate for defining $X_L\times X_R$.)  

In short, the relation between $X_\cs$ and $X_L\times X_R$ is very similar to the relation between $\Sigma$ and $\Sigma_L\times \Sigma_R$.
The main difference is that in the bosonic case, there is a completely natural choice of $\Sigma$ (namely the diagonal in $\Sigma_L\times \Sigma_R$),
but for $X$ a complex supermanifold, it does not seem that there is a completely canonical choice of $X_\cs$.  We can simply take
$X_\cs$ to be any subsupermanifold of $X_L\times X_R$ of codimension $2a|0$ whose reduced space is sufficiently close to
the diagonal in $(X_L\times X_R)_\red$.  

If we wish, we can take the reduced space of $X_\cs$ to be precisely the diagonal in $(X_L\times X_R)_\red$, as we did in section 
\ref{bsc} in the original definition of $X_\cs$.  But even then,
it does not seem that there is a completely canonical choice for $X_\cs$ itself.   

\subsection{Application To String Theory: The Worldsheet}\label{applistring}

The most prominent supermanifolds in superstring perturbation theory are the string worldsheet and the supermoduli space over which one
must integrate to compute scattering amplitudes.  Let us consider both from the viewpoint just proposed.

First we consider the string worldsheet, starting with the heterotic string.  We let $\Sigma_R$ be a super Riemann surface; this is a complex
supermanifold of dimension $1|1$ that obeys some additional conditions that are described, for example, in \cite{supersurf}.
These additional conditions need not concern us here.  We let $\Sigma_L$ be an ordinary Riemann surface that is sufficiently close
to the complex conjugate of the reduced space of $\Sigma_L$.   So $\Sigma_L\times \Sigma_R$ is a complex supermanifold
of dimension $2|1$.  To get a candidate for the worldsheet of a heterotic string, we let $\Sigma$ be a smooth submanifold of $\Sigma_L\times \SIgma_R$
of dimension $2|1$ whose reduced space is sufficiently close to the diagonal in $(\Sigma_L\times \Sigma_R)_\red$.  For example,
we may start by picking $\Sigma_\red\subset (\SIgma_L\times \Sigma_R)_\red$ to be any cycle sufficiently close to the diagonal 
and then ``thicken'' it slightly in the fermionic directions (as explained in section \ref{submanifolds}) to get
$\Sigma$.

If we wish, we can assume that $\Sigma_L$ is precisely the complex conjugate
of $\Sigma_{R,\red}$ and take $\Sigma_\red$ to be the diagonal in $(\SIgma_L\times \Sigma_R)_\red$.  If we do so, then $\Sigma$
can be $\Sigma_{R,\cs}$, the cs version of $\Sigma_R$ as defined (with some choice of gluing law)
in section \ref{bsc}.  But there is no reason to limit ourselves to precisely this case.  At any rate,
even if one wishes to restrict to the case that $\Sigma$ is $\Sigma_{R,\cs}$, the present approach makes it more clear how natural 
$\Sigma_{R,\cs}$ is or is not.

The main thing we need to know to make sure that this approach to heterotic string theory  makes sense is that given any $\Sigma\subset
\Sigma_L\times \Sigma_R $ as above, the worldsheet action of the heterotic string
can be defined as an integral over $\Sigma$, and moreover this integral does not depend on the choice of $\Sigma$.  The relevant facts
are explained in \cite{supersurf}.  
 The Lagrangian density of the heterotic string will be a holomorphic section of
the holomorphic Berezinian  $\mathit{Ber}(\SIgma_L\times \Sigma_R)$ (defined in a suitable neighborhood of the diagonal in the reduced space)
and as explained in section \ref{periods}, this can be integrated over the real cycle $\Sigma$, with a result that only depends on the homology
class of $\Sigma$.  We do not have any way to pick a canonical $\Sigma$, but any choice of $\Sigma$ will lead to the same integrated worldsheet
action.

For Type II superstring theory, the basic idea is the same. The holomorphic and antiholomorphic dimensions of the string worldsheet will
now both be $1|1$, but all the considerations that we have described are still relevant since antiholomorphic odd coordinates are not
supposed to be in any sense complex conjugates of holomorphic odd coordinates.  For Type II superstring theory, we let both $\Sigma_L$
and $\Sigma_R$ be super Riemann surfaces, such that the reduced space of $\Sigma_L$ is sufficiently close to the complex
conjugate of the reduced space of $\Sigma_R$. (No relationship is assumed between the spin structures of $\Sigma_L$ and $\Sigma_R$.)  $\Sigma_L\times \Sigma_R$ is a complex supermanifold of dimension $2|2$.
Now we let $\SIgma$ be any smooth subsupermanifold of $\Sigma_L\times \Sigma_R$ of dimension $2|2$ whose reduced space is sufficiently
close to the diagonal.  Just as for the heterotic string, 
we may start by picking $\Sigma_\red\subset (\SIgma_L\times \Sigma_R)_\red$ to be any cycle sufficiently close to the diagonal 
and then ``thicken'' it slightly in the fermionic directions to get
$\Sigma$.
Again, the worldsheet action can be defined by integrating over $\Sigma$ a holomorphic section of $\mathit{Ber}(\Sigma_L\times \SIgma_R)$
(defined sufficiently close to the diagonal in the reduced space) and the integrated action does not depend on the precise choice of $\Sigma$.

Whenever we say that two objects are ``sufficiently close,'' one may interpret this to mean that the objects in question are equal if
the odd variables are set to zero. (For reduced spaces, this means simply that they are equal.) 
That will always be sufficiently close for any purpose. But in any concrete case, the two objects
need not be quite so close as that, and it is sometimes better to allow oneself a little more elbow room.

\subsection{Application To String Theory: Supermoduli Space}\label{applimoduli}

The other supermanifold that plays a prominent role in superstring perturbation theory is the moduli space over which one integrates
in computing scattering amplitudes.  

In each of the above constructions, let $\M_L$ be the moduli space parametrizing the possible choices of 
$\Sigma_L$, and let $\M_R$ be the moduli space parametrizing $\Sigma_R$.  Concretely, for either the heterotic string 
or Type II, $\M_R$ is the  moduli space $\MM$ of super Riemann surfaces
and its reduced space is what we will call $\M_\spin$, which parametrizes an ordinary Riemann surface with a choice
of spin structure.  As for $\M_L$, for the heterotic string, it is what we will call $\M$, the moduli space of ordinary
Riemann surfaces; for Type II, it is again the moduli space $\MM$ of super Riemann surfaces.  The reduced space
of $\M_L\times \M_R$ is therefore $\M\times \M_\spin$ for the heterotic string, or $\M_\spin\times \M_\spin$ for Type II.

If we simply let $\Sigma_L$ and $\SIgma_R$ vary independently, the moduli space
parametrizing  $\Sigma_L\times \SIgma_R$ is the product $\M_L\times \M_R$.  
However, in string theory, we do not let $\Sigma_L$ and $\Sigma_R$ vary independently.  Roughly speaking,
the bosonic moduli of $\Sigma_L$ are supposed to be complex conjugates of the bosonic moduli of $\SIgma_R$, though we want
no relation between the odd moduli of $\Sigma_L$ and those of $\Sigma_R$.
To implement this, we proceed as follows.  In the reduced space $(\M_L\times \M_R)_\red$,
we define a submanifold $\varGamma_\red$ by requiring that the complex structures of the ordinary Riemann surfaces
parametrized by $\M_{L,\red}$ and $\M_{R,\red}$ are complex conjugates (one assumes no relationship between the spin
structures).  Then in the usual way, we thicken $\varGamma_\red$ slightly to a subsupermanifold $\varGamma\subset \M_L\times 
\M_R$ of the same codimension.\footnote{This definition has been given in \cite{DEF}, p. 95.  As usual, we could replace
$\varGamma_\red$ by any cycle sufficiently close to the one stated.} 

$\varGamma$ has the necessary properties to be the integration cycle of superstring perturbation theory.  The worldsheet path integral determines
a holomorphic section (defined in a neighborhood
of  $\varGamma_\red$) of the holomorphic Berezinian $\mathit{Ber}(\M_L\times \M_R)$.  For the usual reasons, subject to a caveat that we explain momentarily, the integral of this section over $\varGamma$ does not depend on the choice of $\varGamma$.  

The reason that a caveat is needed is that $\M_L$, $\M_R$, and $\varGamma$ are all not compact.  The integrals
that one encounters in superstring perturbation theory have a delicate behavior at infinity.  Dealing with these integrals
requires, among other things, some care in specifying how $\varGamma$ should behave near infinity. The region at infinity
that causes the subtleties is the infrared region, and the subtleties go into showing that superstring perturbation theory has
the expected infrared behavior.

We can summarize all this as follows.
 There are completely natural moduli spaces $\M_L$ and $\M_R$ for antiholomorphic and holomorphic variables.  It seems
 doubtful that there is a natural
 moduli space that combines the two types of variable, but up to homology there is a natural integration cycle $\varGamma\subset \M_L\times \M_R$, which is what
one actually needs for superstring perturbation theory. 
\vskip 1cm
\noindent {\bf Acknowledgments} Research  supported in part by NSF grant  PHY-0969448.  I thank P. Deligne for much valuable advice
over the years on matters related to supergeometry.  Helpful comments on the manuscript have been given by  G. Moore,
T. Voronov,
  P. Zhao, and especially D. Robbins and  B. Safdi.

\bibliographystyle{unsrt}

\end{document}